\documentclass[useAMS,usenatbib,fleqn]{mn2e}
\usepackage{graphicx}
\usepackage{amsmath,amssymb}
\usepackage{bm}
\usepackage{natbib,aas_macros}
\title[Halo stars in the solar neighbourhood]
{Orbital Eccentricity Distribution of Solar-Neighbour Halo Stars}
\author[K. Hattori and Y. Yoshii]
{K. Hattori$^{}$\thanks{E-mail:
khattori@ioa.s.u-tokyo.ac.jp (KH); yoshii@ioa.s.u-tokyo.ac.jp (YY)} and Y. Yoshii$^{}$\footnotemark[1]\\
$^{}$Institute of Astronomy, School of Science, University of Tokyo, 2-21-1, Osawa, Mitaka, Tokyo 181-0015, Japan}

\begin{document}

\date{Accepted 2011 August 15. Received 2011 August 15; in original form 2011 June 10}

\pagerange{\pageref{firstpage}--\pageref{lastpage}} \pubyear{2011}

\maketitle

\label{firstpage}

\begin{abstract}
We present theoretical calculations for the differential distribution of stellar orbital eccentricity for a sample of solar-neighbour halo stars. Two types of static, spherical gravitational potentials are adopted to define the eccentricity $e$ for given energy $E$ and angular momentum $L$, such as an isochrone potential and a Navarro-Frenk-White potential that can serve as two extreme ends covering in-between any realistic potential of the Milky Way halo. The solar-neighbour eccentricity distribution $\Delta N(e)$ is then formulated, based on a static distribution function of the form $f(E,L)$ in which the velocity anisotropy parameter $\beta$ monotonically increases in the radial direction away from the galaxy center, such that $\beta$ is below unity (near isotropic velocity dispersion) in the central region and asymptotically approaches $\sim 1$ (radially anisotropic velocity dispersion) in the far distant region of the halo. We find that $\Delta N(e)$ sensitively depends upon the radial profile of $\beta$, and this sensitivity is used to constrain such profile in comparison with some observational properties of $\Delta N_{\rm obs}(e)$ recently reported by \cite{Carollo2010}. 
Especially, the linear $e$-distribution and the fraction of higher-$e$ stars for their sample of solar-neighbour inner-halo stars rule out a constant profile of $\beta$, contrary to the opposite claim by \cite{Bond2010}. %Bond et al. (2010) %
Our constraint of $\beta \lesssim 0.5$ at the galaxy center indicates that the violent relaxation that has acted on the inner halo is effective within a scale radius of $\sim 10\;{\rm kpc}$ from the galaxy center. We discuss that our result would help understand the formation and evolution of the Milky Way halo. 
\end{abstract}


\section{Introduction} 

Stellar halo of the Milky Way contains invaluable information to understand 
the Galactic formation and evolution, because its collisionless nature has 
preserved some kinematic properties of individual stars in the halo. The most 
commonly probed among them is the orbital eccentricity which remains almost 
unchanged as long as the gravitational potential changes adiabatically. 
This quasi-adiabatic invariance of eccentricity leads to an idea that the 
eccentricity {\it distribution} of halo stars has been conserved until present 
since the end of the last rapid change of the potential. Such rapid change may 
be related to some dynamical events -- such as the major merger -- in the 
course of forming the Milky Way and may have caused the violent relaxation 
of the stellar system (\citealt{Lynden-Bell1967}). Thus, detailed analysis of 
the observed eccentricity distribution of halo stars would provide us with a 
clue for looking into the early dynamical state of the Milky Way. 

The first major milestone in the statistical study of orbital eccentricity of 
halo stars was made by Eggen, Lynden-Bell \& Sandage (1962), %\cite{ELS}, 
who used a sample of 221 solar-neighbour 
stars with available $(U,V,W)$-velocity data. They calculated the model 
orbits of individual stars and found a correlation between ultraviolet excess 
and orbital eccentricity, such that metal-poor stars, which constitute the 
stellar halo, are in eccentric orbits with complete lack of nearly circular 
orbits. They interpreted this as the evidence that the stellar halo of the 
Milky Way formed out of a rapidly collapsing proto-Galactic gas cloud. However, 
their sample of halo stars, mostly taken from the catalog of high proper motions, 
turned out to be biased toward eccentric orbits. In fact, based on non-kinematically selected samples, 
various authors did show the existence of halo stars in nearly circular orbits and required the original interpretation of the halo collapse 
by \cite{ELS} to be significantly modified 
(\citealt{Yoshii1979}; Norris, Bessell, \& Pickles 1985; %\citealt{Norris1985}; 
\citealt{Chiba1997}).

Some attempts have been made along to explain the observed eccentricity distribution 
of halo stars. For example, \cite{Chiba1998} showed that a simple form of the distribution function that allows a radially-anisotropic velocity dispersion is more or less consistent with the eccentricity distribution obtained from the Hipparcos data. However, because of the small sample size, they could not go further to fully interpret the eccentricity distribution, i.e., they could not well constrain the global dynamical structure over the halo by using the eccentric orbits that could travel far beyond the solar neighbourhood.

Recently, with advent of huge surveys including the SDSS (Sloan Digital Sky Survey), much larger samples of halo stars are available, and it has become possible to 
discuss the observed eccentricity distribution on a firm statistical basis. 
In particular, \cite{Carollo2010} analysed $\sim10,000$ SDSS spectra of 
non-kinematically selected calibration stars, and derived their metallicities and radial velocities with reasonable accuracy.  Using these as well as the proper motions from other catalogs, they found a marked metallicity dependence in the eccentricity distributions of halo stars obtained in different ranges of metallicity. They argued that this metallicity dependence would have arisen from a distance-dependent mixture ratio of two distinct stellar components of inner and outer halos having different eccentricity distributions and different metallicities.  
While their argument is intriguing in itself, more theoretical works are necessary 
to have insight into the origin of the dichotomic halo. 

As a first step toward improving current understanding of the eccentricity distribution, \citeauthor{Hattori2010} (2010; HY hereafter) performed theoretical calculations of eccentricity distribution for the {\it whole} halo by modeling 
the gravitational potential and the distribution function. When applying to the 
{\it solar-neighbour} halo stars, however, the analysis in HY has to be modified, because such a sample is biased against stars that spend less fraction of orbital period in the solar neighbourhood. In this paper, by explicitly taking this bias 
into account, we re-formulate the analysis in HY and properly compare our 
theoretical distribution to that observed for the solar-neighbour halo 
stars. In this way, we would be able to constrain the halo dynamics hopefully 
back to an epoch of violent relaxation that occurred in forming the Milky Way. 

In section 2, we present our new formulation for theoretical eccentricity 
distribution of the solar-neighbour halo stars $\Delta N(e)$, with an 
emphasis on the difference from the formulation in HY. In section 3, we calculate 
$\Delta N(e)$ for two spherical models of the Milky Way halo. In section 4, 
we demonstrate how our result in section 3 is beneficial to interpret 
the solar-neighbour data. In section 5, we summarize the basic ideas from our 
analysis.

\section{Formulation} \label{bias}

\subsection{Survey region} 

As in HY, we assume that the stellar halo is a spherical system that consists of many halo stars in a steady, spherical gravitational potential of the dark matter halo. We further assume that kinematic data of halo stars can be obtained only in a shell-like region defined as 
\begin{equation} \label{survey region}
r_{-} < r < r_{+} , 
\end{equation} 
where $r\equiv|{\bm r}|$ is the Galactocentric distance. We shall hereafter call this the survey region. 
We note that the actual survey region in this shell should be confined in the solar neighbourhood, 
centered at the position of the sun. 
However, this realistic definition of the actual survey region barely makes any difference 
to any result from the shell-like region [inequality (\ref{survey region})] 
as far as the halo is assumed to be of spherical symmetry in our analysis. 
See also Appendix \ref{Lindblad} and footnote \ref{fn}.

\subsection{Stellar orbital eccentricity in a model halo} \label{e}

Given a spherical halo potential $V(r)$, the orbital eccentricity of a star is practically defined as 
\begin{equation} \label{e-def}
e \equiv \frac{ r_{\rm apo} - r_{\rm peri} }{ r_{\rm apo} + r_{\rm peri} } ,
\end{equation}
where $r_{\rm apo}$ and $r_{\rm peri}$ are the apocenter and pericenter distances, respectively, and are given by two real solutions ($r_{\rm apo}>r_{\rm peri}$) of the following equation: 
\begin{equation} \label{eq_r1r2}
E = V(r) + \frac{L^2}{2 r^2} \equiv V_{\rm{eff}} (L;r) , 
\end{equation}
where $E$ and $L$ are the specific energy and angular momentum, respectively. 
As discussed in Appendix A of HY, bound stars ($E<0$), 
provided they are observable in the survey region, 
reside only in a limited region of the ($E,L$)-phase space, and we shall concentrate on these stars below. 

\subsection{Differential eccentricity distribution of observable stars} \label{formulation}

The distribution function for a system of spherical symmetry takes the form $f(E,L)$, according to the strong Jeans theorem (\citealt{Lynden-Bell1960}; \citealt{Lynden-Bell1962}). By changing variables and integrating over spherical coordinates, we obtain the number of stars in the survey region [inequality (\ref{survey region})] and in a phase space volume $dE dL^2$ at an epoch of observation: 
\begin{equation}
\Delta N(E,L)\;dE dL^2 = 4 \pi^2 f(E,L) \Delta T_r(E,L) \; dE dL^2. 
\end{equation}
This equation is essentially the same as equation (7) of HY, except that the radial period $T_r$ in HY 
is replaced by the {\it observable time} $\Delta T_r$ 
which is the total length of time per radial period a star with a given $E$ and $L$ spends inside the survey region: 
\begin{multline}\label{DTr} %\begin{equation} 
\Delta T_r (E,L) \equiv \\
2 \theta(\min (r_+,r_{\rm apo}) - \max (r_-,r_{\rm peri})) 
\int_{\max (r_-,r_{\rm peri})}^{\min (r_+,r_{\rm apo})} \frac{dr}{|v_r|} , 
\end{multline}%\end{equation}
where $\theta (x)$ is Heaviside step function. Note that observable stars that satisfy 
\begin{equation}
r_{-} < r_{\rm apo} \;\; \text{and} \;\;  r_{\rm peri} < r_{+} , 
\end{equation}
pass through the survey region at an epoch of observation with a probability $\Delta T_r / T_r$. In particular, observable stars with 
\begin{equation}
r_{-} < r_{\rm peri} < r_{\rm apo} < r_{+} 
\end{equation} 
are always inside the survey region and $\Delta T_r$ coincides with $T_r$. 
On the other hand, unobservable stars, for which 
\begin{equation}
r_{\rm apo} < r_{-} \;\; \text{or} \;\; r_{+} < r_{\rm peri} , 
\end{equation} 
do not pass through the survey region and therefore $\Delta T_r = 0$, 
as clearly seen from equation (\ref{DTr}). 

By using these quantities and remembering that $L^2$ is a function of $E$ and $e$, we obtain the $E$-dependent differential eccentricity distribution of halo stars in the survey region [inequality (\ref{survey region})]:
\begin{equation} \label{DeltanEe}
	\Delta n(E,e) = 4 \pi^2 f(E,L) \Delta T_r(E,L) \left|\left(\frac{\partial L^2}{\partial e}\right)_{E}\right|  . 
\end{equation}
This expression differs from HY, i.e., we have included the distribution function $f(E,L)$, 
instead of placing it outside as a weight to be added to 
the $E$-dependent eccentricity distribution in equation (8) of HY.  
We then express the differential eccentricity distribution as
\begin{equation} \label{DeltaNe}
\Delta N(e) 
= \int_{V(r_{-})}^{0} \Delta n(E,e) \; dE . 
\end{equation}
Here, the integral over $E$ should be performed over the energy range of bound, observable stars with $\Delta T_r > 0$. The allowed $E$-region for such stars is 
presented in the $(E,L)$ diagram in Appendix \ref{Lindblad}.

\subsection{Distribution function of stellar halo}

Our knowledge of the distribution function of the stellar halo $f(E,L)$ is fragmentary. 
Observationally, it is indicated that the first moments or the mean values of velocity components are near zero except for the rotation component (e.g. \citealt{Chiba2000}). [In fact, the first moments for the spherical model with no systematic rotation are zero.] Thus, the most useful constraint on $f(E,L)$ is the behaviour of the velocity anisotropy which is usually parameterised as
\begin{equation}
\beta \equiv 1 - \frac{\sigma_{\rm t}^2}{\sigma_r^2}, 
\end{equation}
where $\sigma_{r}$ is the radial velocity dispersion and $\sigma_{\rm t}$ is the tangential velocity dispersion projected onto the spherical $\theta$-$\phi$ surface. 

Estimates of $\beta$ in the solar neighbourhood are rather convergent 
and are in a range of $\beta_{\odot}=0.4-0.7$ 
(e.g., \citealt{Yoshii1979}; Gilmore, Wyse \& Kuijken 1989; %\citealt{Gilmore1989}; 
\citealt{Carollo2010}), 
while its value beyond the solar circle is less certain. 
\cite{SL1997} used a sample of field horizontal branch stars and claimed that 
$\beta$ decreases with increasing $r$ 
(i.e., the tangential anisotropy in velocity dispersions dominates in the far distant halo), 
which they regarded as a sign for the bottom-up galaxy formation scenario, 
while \cite{Bond2010} 
used a stellar sample of SDSS and suggested that 
$\beta$ is more or less constant within $5\;{\rm kpc}$ of the sun. 
On the other hand, 
$N$-body 
simulations suggest that the cold collapse of the halo, 
which is most likely to trigger the violent relaxation, 
favours $\beta$ that increases with increasing $r$ (e.g. \citealt{vanAlbada1982}; \citealt{Voglis1994}). 
Similarly, recent $N$-body + gas-dynamical simulations suggest that 
$\beta$ of a stellar halo increases with increasing $r$ 
if it is formed through disruption of satellite galaxies. 
(e.g. Abadi, Navarro \& Steinmetz 2006).

In this paper, following \cite{Cuddeford1991}, 
we model the stellar halo as a family of distribution functions of the form:\footnote{
The allowed region of the $(E,L)$-phase space for bound stars is described in Appendix A of HY. 
} 
\begin{equation}\label{DFB}
	f(E,L) = 
	\begin{cases}
		g(Q) L^{-2 \beta_0} , & \text{if $(E,L)$ is `allowed'} \\
		0 ,                      & \text{\rm otherwise,} 
	\end{cases}
\end{equation}
where $g(Q)$ is an arbitrary function of $Q$ which is a linear combination of $E$ and $L^2$ [see section \ref{dfn} below], 
and $\beta_0$ is $\beta$ at the galaxy center.  
In this model, the radial profile of $\beta$ as a function of $r$ is expressed as 
\begin{equation} \label{betar}
	\beta = \beta_0 \frac{r_a^2}{r^2 + r_a^2} + \frac{r^2}{r^2 + r_a^2} , 
\end{equation}
where $r_a$ is the so-called anisotropy radius which characterizes the profile of $\beta$. 
[In section \ref{bias_halo}, we re-arrange this form in terms of $\beta_0$ and $\beta_{\odot}$ instead of $\beta_0$ and $r_a$.] 

The profile of $\beta$ for $r_a=\infty$ reduces to a constant profile with $\beta = \beta_0$. 
On the other hand, the profile for a finite value of $r_a$ behaves such that 
$\beta$ increases monotonically from $\beta_0$ at $r=0$ to $\beta=1$ at $r \gg r_a$. 
The functional form of the profile in equation (\ref{betar}), though not covering $\beta$ that decreases with increasing $r$, 
is enough to examine effects of the radial gradient in $\beta$ upon $\Delta n(E,e)$ as well as $\Delta N(e)$. 
In a forthcoming paper (Hattori 2011, in preparation), we will discuss 
another family of distribution functions in which $\beta_0 > \beta_{\odot}$ is also allowed.

\subsection{Limiting cases of the survey region} 

As mentioned above, the finite radial thickness $|r_{+}-r_{-}|$ of the survey region 
affects directly $\Delta N(e)$, 
but the difference shows up only through $\Delta T_r$. 
To see further this effect, it is instructive to consider two limiting cases: 
(1) the large-shell limit of $r_{-}=0$ and $r_{+}\to \infty$, and 
(2) the thin-shell limit of $r_{-}=r_{\odot}-\delta r/2$ and $r_{+}=r_{\odot}+\delta r/2$ with $\delta r \to 0$, 
where $r_{\odot}$ is the radial distance to an observer from the galaxy center. 

Obviously, in the case (1), it follows that $\Delta T_r =T_r$ and the result of $\Delta N(e)$ is the same as $N(e)$ given in HY.  Therefore, in this subsection, we do not repeat the result of this case. In the case (2) of thin-shell limit, use of equation (\ref{DTr}) gives 
\begin{equation}\label{thinDTr}
\Delta T_r \to 2 \frac{\delta r}{v_{r \odot}} 
\theta(r_{\rm apo} - r_{\odot}) \theta(r_{\odot}-r_{\rm peri}) , 
\end{equation}
where $v_{r \odot}=|v_r|$ is the radial speed at $r=r_{\odot}$ which depends on $E$ and $e$.\footnote{ 
Strictly speaking, equation (\ref{thinDTr}) does not hold when $v_{r \odot} = 0$. 
However, we neglect such a physically trivial exception in our analysis.} 
In the following discussion, we consider the eccentricity distribution, 
using a distribution function of the above form [equation (\ref{DFB})] 
with a constant profile of $\beta$ ($r_a=\infty$) in two extreme gravitational potentials of the central point mass 
and the truncated homogeneous density distribution. We note that these potentials allow analytic expression of $\Delta n_{\beta}(E,e)$ and serve as useful reference. 
Here and hereafter, the subscript $\beta$ to $\Delta n(E,e)$ or $\Delta N(e)$ stands for the case of constant profile of $\beta$. 
For our explanation below, 
it is useful to introduce the specific energy of a star in a circular orbit with orbital radius $r$: 
\begin{equation} \label{Ec}
E_{\rm c}(r) \equiv V(r) + \frac{1}{2}r\frac{d}{dr}V(r). 
\end{equation}

\subsubsection{Central point mass}\label{Kep}

The gravitational potential arising from the central point mass is Keplerian 
and is given by $V(r) =  - GM/r$, where $M$ is the mass and $G$ is the gravitational constant. 
By evaluating $r_{\rm apo}$ and $r_{\rm peri}$, we obtain
\begin{equation}\label{theta_Kep}
	\theta (r_{\rm apo} - r_{\odot}) \theta (r_{\odot} - r_{\rm peri}) 
	= \theta \left( e - \left|\frac{E}{E_{\odot}} -1 \right| \right) , 
\end{equation}
where $E_{\odot} = E_{\rm c}(r_{\odot}) = -GM/(2 r_{\odot})$.

When $E<V(r_{\odot})$, equation (\ref{theta_Kep}) vanishes, therefore $\Delta T_r = 0$ and $\Delta n_{\beta} = 0$. On the other hand, when $V(r_{\odot})<E<0$, $\Delta T_r$ can be expressed as 
\begin{equation} \label{DTr_Kep}
	\Delta T_r = \frac{4 \delta r}{\sqrt{-2E}} \frac{\theta (e - e_{\rm cut})}{ \sqrt{e^2 - e_{\rm cut}^2} } , 
\end{equation}
which allows the analytic expression of $\Delta n_{\beta}(E,e)$ for a constant profile of $\beta$ as follows:  
\begin{multline} \label{Delta_nA_Kep}
	\Delta n_{\beta}(E,e) = \\
	32 \pi^2 \delta r {(GM)}^{2-2\beta} 
	\frac{g(E)}{{(-2E)}^{3/2}} 
	\frac{e}{{(1-e^2)}^{\beta}}  \frac{\theta (e - e_{\rm cut})}{\sqrt{e^2 - e_{\rm cut}^2}} , 
\end{multline}
where the cutoff eccentricity $e_{\rm cut}$ is defined as 
\begin{equation} \label{cutoff_Kep}
       e_{\rm cut} \equiv |E/E_{\odot} -1| . 
\end{equation}
Evidently, $e_{\rm cut}$ gives the same value for a pair of two energies $E$ and $W$ which are related to each other via $E-E_{\odot} = E_{\odot}-W$. 
Therefore, we see from equation (\ref{Delta_nA_Kep}) that the shape of $\Delta n_{\beta}(E,e)$ 
for a given $E$ is identical with that of $\Delta n_{\beta}(W,e)$ for the corresponding $W$. 
Because of this relation, 
we only need to know the shape of $\Delta n_{\beta}(E,e)$ for either $E \geq E_{\odot}$ or $E \leq E_{\odot}$. 
Results of $\Delta n_{\beta}(E,e)$ for 
several combinations of $\beta$ and $E$ 
are shown on the left-hand panels of Figure \ref{DeltanEeKepIHO}.

%%%%%%%%%%%%%%%%%%%%%%%%%%%%%%%%%%%%%%%%%%%%%% IHO  homogeneous  
\subsubsection{Truncated homogeneous sphere}\label{IHO}

A homogeneous density distribution within truncated sphere is expressed as    
$\rho(r) = {3M}/(4 \pi r_{\rm t}^3)$ at $r < r_{\rm t}$ and $\rho(r)=0$ at $r>r_{\rm t}$, where $M$ is the total mass of dark halo and $r_{\rm t}$ is the truncation radius. The gravitational potential arising from this density distribution is given by  $V(r) = E_{\rm min} + GMr^2/(2 r_{\rm t}^3)$ at $r < r_{\rm t}$, where $E_{\rm min}=-3GM/(2r_{\rm t})$. If the stellar system is confined inside the truncation radius $r_{\rm t}$, by evaluating $r_{\rm apo}$ and $r_{\rm peri}$, we obtain 
\begin{equation}\label{theta_IHO}
	\theta (r_{\rm apo} - r_{\odot}) \theta (r_{\odot} - r_{\rm peri}) 
	= \theta \left( \frac{2e}{1+e^2} - \frac{ \left| E - E_{\odot} \right| }{E-E_{\rm min}} \right) , 
\end{equation}
where $E_{\odot} = E_{\rm c}(r_{\odot}) = E_{\rm min} + GMr_{\odot}^2/r_{\rm t}^3$.

When $E<V(r_{\odot})$, equation (\ref{theta_IHO}) vanishes, and therefore $\Delta T_r = 0$ and $\Delta n_{\beta} = 0$. On the other hand, when $V(r_{\odot})<E<V(r_{\rm t})$, equation (\ref{theta_IHO}) reduces to $\theta (e -e_{\rm cut})$ and thus $\Delta T_r$ can be expressed as 
\begin{equation}
	\Delta T_r = \frac{2 \delta r}{\sqrt{GM/r_{\rm t}}} \left(\frac{r_{\rm t}}{r_{\odot}}\right) 
	\frac{E-E_{\odot}}{E-E_{\rm min}}  
	\frac{\theta(e - e_{\rm cut})}
	{\sqrt{ {\left( \frac{2e}{1+e^2} \right)}^2 - {\left( \frac{2 e_{\rm cut}}{ 1 + e_{\rm cut}^2 } \right)}^2 } }, 
\end{equation}
which allows the analytic expression of $\Delta n_{\beta}(E,e)$ for a constant profile of $\beta$ as follows:  
\begin{multline}\label{Delta_nA_IHO}
	\Delta n_{\beta}(E,e) = \\
	64 \pi^2 \frac{\delta r}{r_{\odot}} {\left( \frac{r_{\rm t}^3}{GM} \right)}^{3/2 - \beta} 
	(E_{\odot}-E_{\rm min}) (E-E_{\rm min})^{1-2\beta} g(E) \\
	\times
	\frac{e {(1-e^2)}^{1-2\beta}}{{(1+e^2)}^{3-2\beta}} 
	\frac{\theta(e - e_{\rm cut})}
	{\sqrt{ {\left( \frac{2e}{1+e^2} \right)}^2 - {\left( \frac{2 e_{\rm cut} }{ 1 + e_{\rm cut}^2 } \right)}^2 } },
\end{multline} 
where the cutoff eccentricity $e_{\rm cut}$ is defined as  
\begin{equation}\label{cutoff_IHO}
e_{\rm cut} \equiv \frac{E - E_{\rm min}}{|E - E_{\odot}|} 
- \sqrt{ {\left( \frac{E-E_{\rm min}}{E-E_{\odot}} \right)}^2 - 1 }. 
\end{equation} 
Similarly to the case of central point mass, $e_{\rm cut}$ gives the same value for a pair of two energies $E$ and $W$ 
which are related to each other via $(E - E_{\rm min})/(E - E_{\odot})=(W - E_{\rm min})/(E_{\odot}-W)$. 
Therefore, we see from equation (\ref{Delta_nA_IHO}) that the shape of $\Delta n_{\beta}(E,e)$ 
for a given $E$ is identical with that of $\Delta n_{\beta}(W,e)$ for the corresponding $W$. 
Because of this relation, 
we only need to know the shape of $\Delta n_{\beta}(E,e)$ for either $E \geq E_{\odot}$ or $E \leq E_{\odot}$, 
although we have to keep in mind that we truncate the stellar system at $E=V(r_{\rm t})$. 
Results of $\Delta n_{\beta}(E,e)$ with $r_{\odot}=0.15r_{\rm t}$ for 
several combinations of $\beta$ and $E$ 
are shown on the right-hand panels of Figure \ref{DeltanEeKepIHO}.

\subsubsection{Insights from the thin-shell limit}

In HY, $n_{\beta}(E,e)$ for the {\it whole} stellar system with a constant profile of $\beta$ is found to be given by $F(E)\times G(e)$ in the extreme gravitational potentials considered here, so that the shape of $n_{\beta}(E,e)$ does not depend on $E$ (see Figure 1 of HY). 
On the other hand, $\Delta n_{\beta}(E,e)$ for the survey region described above depends on $E$. Since $\Delta n_{\beta} / n_{\beta} \propto  \Delta T_r / T_r$ \footnote{
Notice that a way of defining $n_{\beta}$ without $g(E)$ as in equation (8) of HY differs from $\Delta n_{\beta}$ which includes $g(Q)$ as in equations (\ref{DeltanEe}) of this paper.
} 
and $T_r$ does not depend on $e$,\footnote{
This is the case for these extreme potentials considered here as well as in the isochrone potential below [see equations (13), (23), and (33) of HY]. 
In general, $T_r$ depends on both $E$ and $e$, but $e$-dependence is often very weak. 
}
extra $e$-dependence of $\Delta n_{\beta}$ over $n_{\beta}$ is attributed exclusively to the $e$-dependence of $\Delta T_r$. 

In the thin-shell limit, $\Delta T_r$ shows a marked $e$-dependence, especially near $e=e_{\rm cut}$: it vanishes at $0 \leq e \leq e_{\rm cut}$, 
diverges at $e=e_{\rm cut}+0$, and rapidly decreases at $e>e_{\rm cut}$, 
with its slope becoming less and less steep toward higher $e$. This marked $e$-dependence of $\Delta T_r$ is reflected to $\Delta n_{\beta}$ and the shape of $\Delta n_{\beta}$ at $e \lesssim e_{\rm cut}$ is essentially determined by $\Delta T_r$, regardless of $\beta$. At higher-$e$ region, however, the $L^{-2\beta}$ term in the distribution function [equation (\ref{DFB})] 
starts to affect $\Delta n_{\beta}$ with its slope continuing to decrease toward higher $e$ for $\beta \lesssim 0.5$, while showing upturn there for $\beta \gtrsim 0.5$. 

\begin{figure*}
	\begin{center}
	\includegraphics[width=\columnwidth]{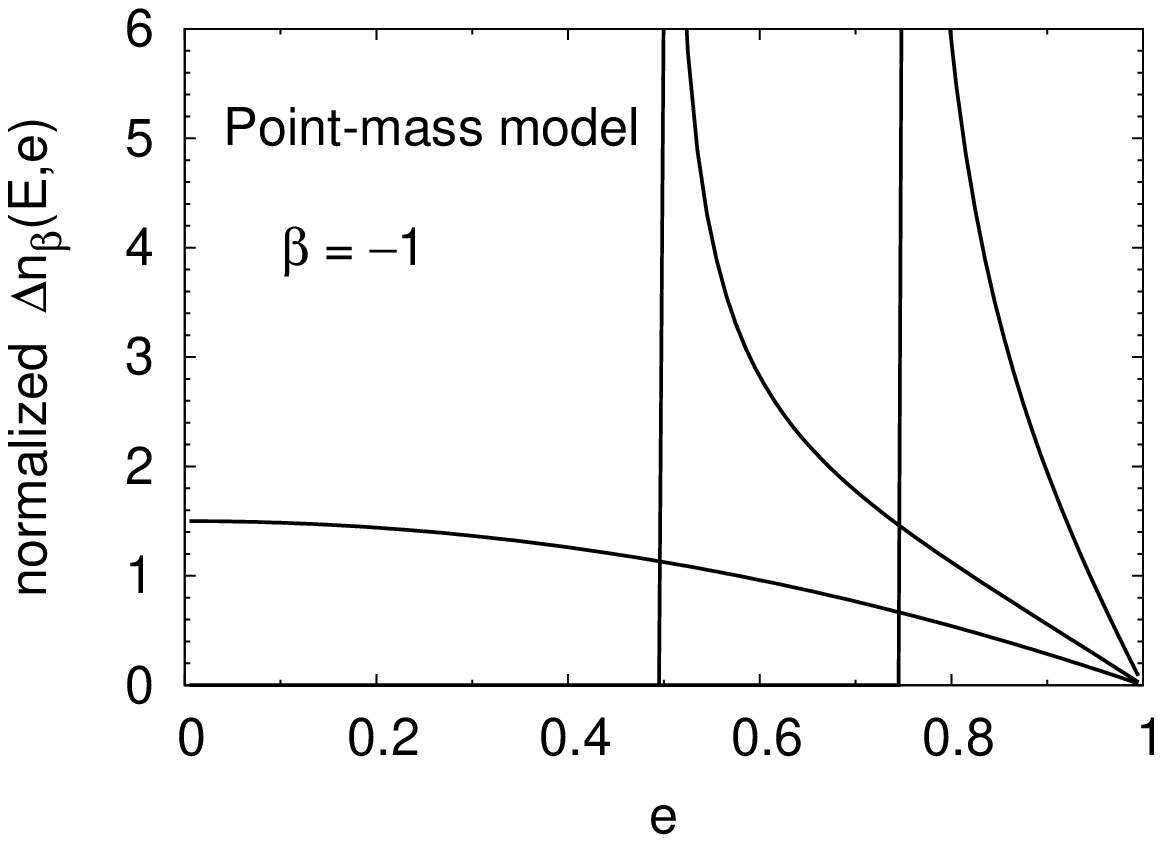}
	\includegraphics[width=\columnwidth]{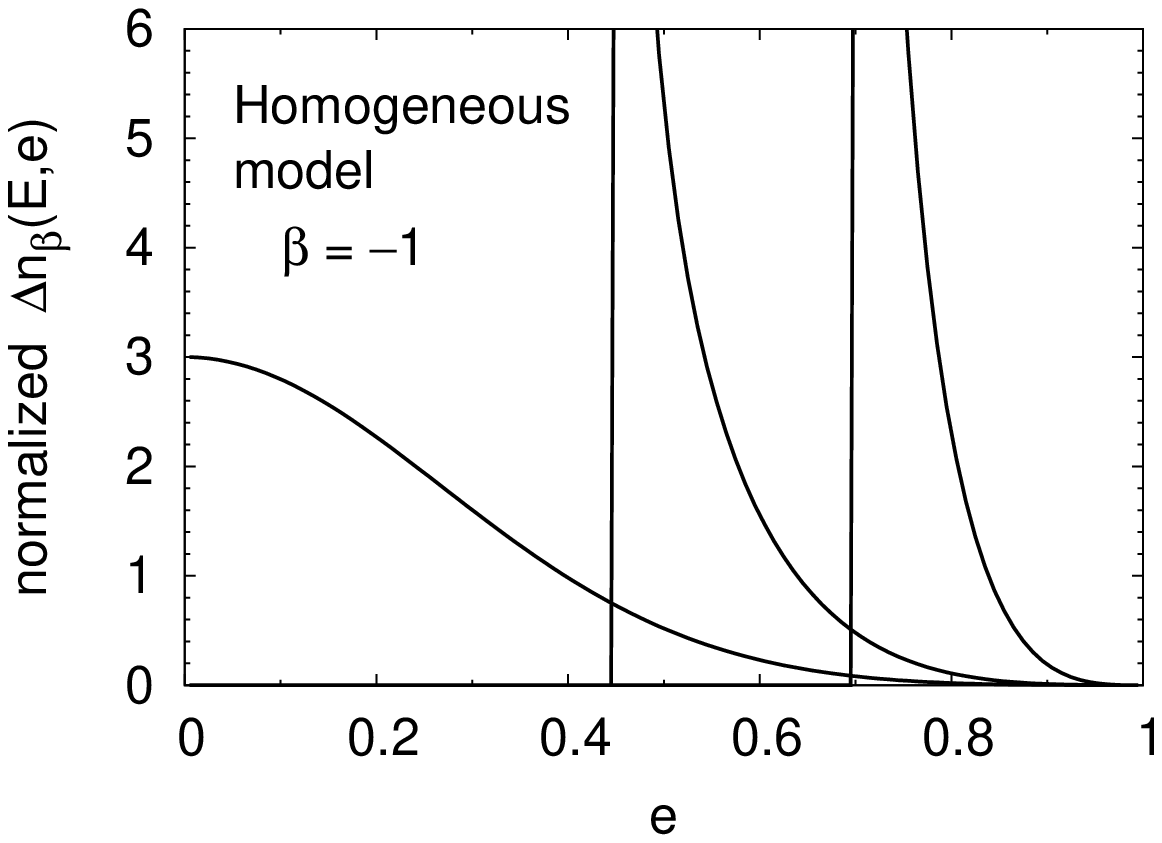}\\
	\includegraphics[width=\columnwidth]{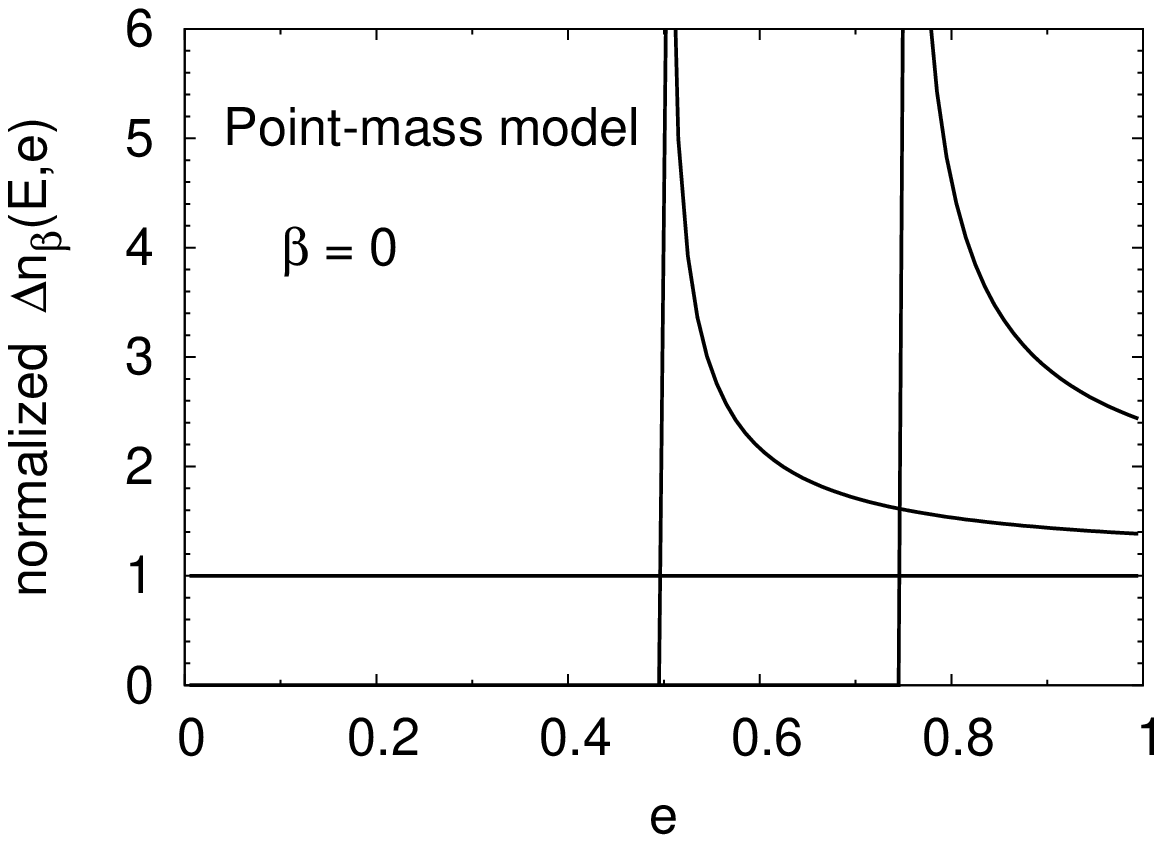}
	\includegraphics[width=\columnwidth]{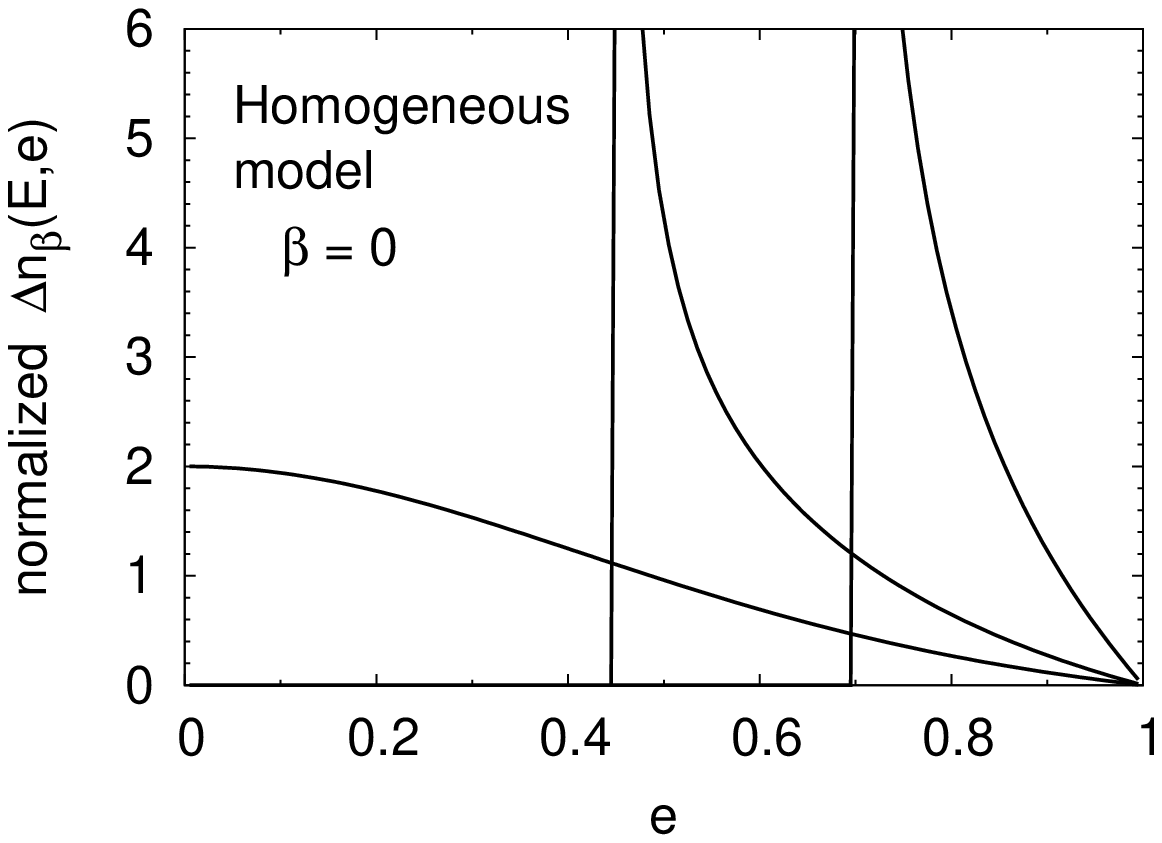}\\
	\includegraphics[width=\columnwidth]{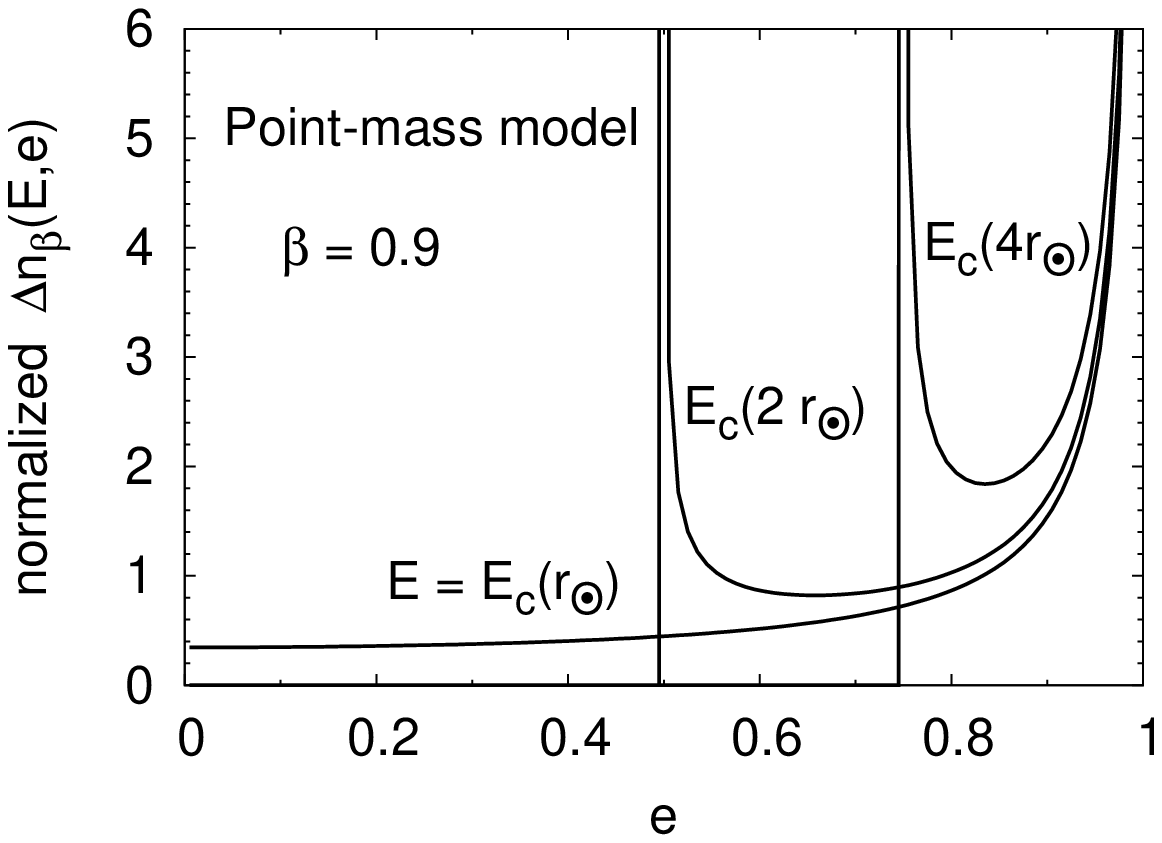}
	\includegraphics[width=\columnwidth]{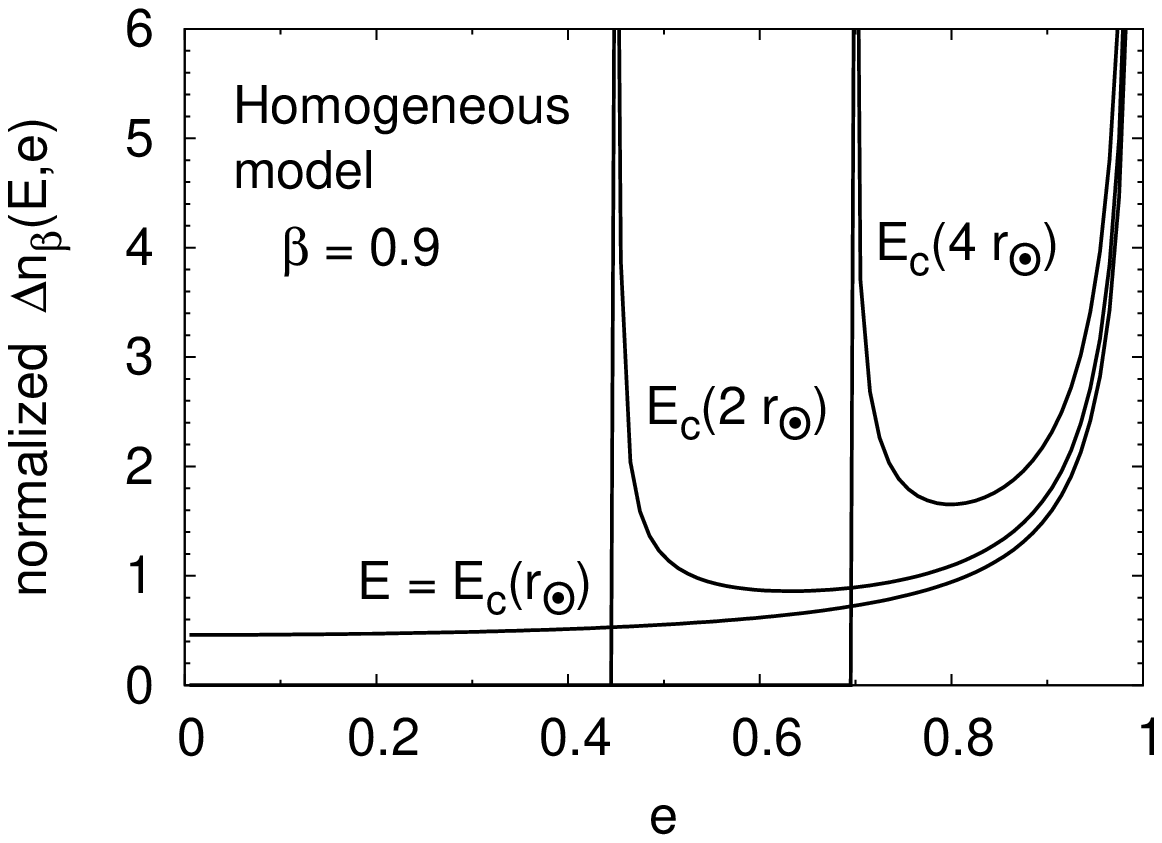}\\
	\caption{
Energy-dependent differential distribution $\Delta n_{\beta}(E,e)$ of stellar orbital eccentricity 
in a spherical system with constant radial $\beta$-profile for an observer at $r_{\odot}=8.5\;{\rm kpc}$ 
whose survey region is limited to nearby stars only (thin-shell limit). 
Shown are the results in two extreme models of gravitational potential, 
such as the point-mass model for several values of $\beta$ on the left-hand panels (section \ref{Kep}) 
and the homogeneous model with $r_{\odot}/r_{\rm t}=0.15$ for several values of $\beta$ on the right-hand panels (section \ref{IHO}). 
Shown by lines on each panel for given $\beta$ are the results for several values of specific energy $E$ 
equated to $E_c(r)$ of a star in circular orbit at different orbital radius $r$. 
Note that $\Delta n_{\beta}(E,e)$ is normalized such that $\int_0^1 \Delta n_{\beta}(E,e) de = 1$. 
} 
\label{DeltanEeKepIHO}
	\end{center}
\end{figure*}

In section 2.4 of HY, we have shown 
that $n_{\beta}$ for more centrally concentrated system is more weighted toward high-$e$ region. 
This trend can be confirmed for $\Delta n_{\beta}$ shown in Figure \ref{DeltanEeKepIHO}. 
We see from this figure that for given $\beta$ and $E$, 
$\Delta n_{\beta}$ for the central point-mass model is more weighted toward high-$e$ region than 
that for the truncated homogeneous sphere model. 
However, Figure \ref{DeltanEeKepIHO} also suggests that this trend is not very prominent for a stellar system with $\beta \sim 1$.

From the analytic expression of $e_{\rm cut}$, we see that $\Delta n_{\beta}$ depends 
not only on specific energy $E$ of observed stars\footnote{
For the $E$-dependence of $e_{\rm cut}$, see Appendix \ref{Lindblad}.} 
but also on $E_{\odot}$ which is a function of $M$ and $r_\odot$ for the case of central point mass, 
or a function of $M$, $r_\odot$, and $r_{\rm t}$ for the case of truncated homogeneous sphere. 
Thus, realistic modeling of the Milky Way is necessary to perform theoretical calculations of $\Delta n_{\beta}$ expected at the position of the sun.

% Kepler+IHO bias   

\section{Eccentricity distribution of solar-neighbour halo stars} \label{bias_halo}

The formulation in the previous section can be applied to more general 
gravitational potentials, such as the isochrone and NFW potentials, 
which we consider to be limiting cases of reasonable descriptions of the mass 
distribution in dark haloes. 
In this section, 
we derive $\Delta N(e)$ for these potentials by specifying the survey region and the distribution function.

\subsection{Calculation setup} \label{setup}

\subsubsection{Gravitational potential} \label{pot}

The gravitational potential of the isochrone model (\citealt{Henon1959}) 
is given by $V(r) = - GM/(b + \sqrt{b^2 + r^2})$, where $M$ is the total mass and $b$ is the scale length parameter. For the Milky Way halo, following \cite{Chiba1998}, we set $b=5.2\; {\rm kpc}$ and $\sqrt{GM/(2b)}=385\;{\rm km\;s^{-1}}$.

The NFW density profile of dark matter produces a gravitational potential of the form 
$V(r) = -4\pi G {\rho}_0 a^3 [\ln(1+ r/a)]/r$, 
where $\rho_0$ is the typical density and $a$ is the scale length parameter (Navarro, Frenk \& White 1997). 
Considering typical values for the Milky Way-size dark matter haloes in recent cosmological simulations (e.g., \citealt{Zhao2003}), we set $a=20\;{\rm kpc}$ and $\sqrt{4\pi G\rho_0 a^2}=360\;{\rm km\;s^{-1}}$, together with additional values of $r_{200}=200\;{\rm kpc}$ for the virial radius and $c\equiv r_{200}/a=10$ for the concentration parameter.

\subsubsection{Survey region}

The survey region adopted is given by $r_{-}=7\;{\rm kpc}$ and $r_{+}=10\;{\rm kpc}$ in inequality (\ref{survey region}). This region matches with the spatial criteria for sampling halo stars by \cite{Carollo2010}, namely, $7\;{\rm kpc}<R<10\;{\rm kpc}$ ($R$ is the projected Galactocentric distance onto the disk plane) and $d<4\;{\rm kpc}$ ($d$ is the distance from the sun).

\subsubsection{Distribution function} \label{dfn}

The distribution function adopted is given by the form in equation (\ref{DFB}), with radial profile of $\beta$ which is re-parameterised in terms of $\beta_0$ at the center and  $\beta_\odot$ at $r_\odot$, instead of $\beta_0$ and $r_a$ in equation (\ref{betar}). In our analysis below, we consider several combinations of ($\beta_0, \beta_\odot$) provided $0 \leq \beta_0 \leq \beta_\odot$. In particular, for $\beta_0=\beta_\odot$, a constant profile results.

As for $g(Q)$ with the argument defined as $Q \equiv E+ L^2/(2 r_a^2)$, we adopt 
\begin{equation}\label{gQ}
	g(Q) = 
	\begin{cases}
		A \left( \exp \left[ - \frac{Q-E_{\rm t}}{\sigma^2} \right] - 1 \right), 
		& \text{if $Q<E_{\rm t}$} \\
		0 , 
		& \text{\rm otherwise,} 
	\end{cases}
\end{equation}
where 
$A$ is a constant, $E_{\rm t}$ is a truncation energy, $\sigma$ is a velocity parameter described below.

In general, 
truncation of the distribution function at $Q=E_{\rm t}$ 
for a stellar system in a spherical potential $V(r)$ 
guarantees stars to be confined within a truncation radius $r_{\rm t}$ for which $V(r_{\rm t})=E_{\rm t}$.\footnote{
This is because $r_{\rm apo}$ of a star increases with $E$ when $L^2$ is fixed, while it decreases with $L^2$ when $E$ is fixed.
} 
For the isochrone potential, we express the dimensionless truncation energy as $\varepsilon_{\rm t} \equiv 2bE_{\rm t}/(GM)$. 
Using $r_{\rm t}=200\;{\rm kpc}$ as a characteristic value of reference, we obtain $\varepsilon_{\rm t}= -0.05$. 
On the other hand, for the NFW potential, as in HY, 
we express it as $\varepsilon_{\rm t} \equiv E_{\rm t}/(4\pi G \rho_0 a^2) = - [\ln (1 + c)]/c$, 
by identifying the virial radius $r_{200}$ with the boundary of dark matter halo. 
Then, using $c=10$ (section \ref{pot}), we obtain $\varepsilon_{\rm t}=-0.23$.  

%%%%%
In our model, the radial velocity dispersion $\sigma_r$ decreases with increasing $r$, 
which is consistent with some observations (e.g. \citealt{Brown2010}). 
In addition, under a fixed set of the gravitational potential and the truncation energy $E_{\rm t}$, 
the radial profile of $\sigma_r / \sigma$ is determined by $\beta_0$ only\footnote{
The radial profile of $\sigma_{\rm t} / \sigma$ is determined by $r_a$ as well as $\beta_0$, 
since $\sigma_{\rm t}^2 / \sigma_r^2 = (1-\beta_0) r_a^2 / (r_a^2 + r^2)$. 
} 
and this $\beta_0$-dependence is fairly weak.\footnote{
For $-1/2<\beta_0<1/2$, $\sigma_r / \sigma$ is different from that for $\beta_0=0$ 
by only 8\% (relative difference). 
This percentage becomes no larger than 19\% for $-1<\beta_0<1$. 
} 
In this paper, $\sigma$ is set to be $200\;{\rm km\;s^{-1}}$ so that 
$\sigma_r$ decreases with $r$ from the central region, reaches $\simeq 160\;{\rm km\;s^{-1}}$ in the solar neighborhood, 
and vanishes at $r=r_{\rm t}$.

\subsection{Results} 

\subsubsection{Energy-dependent eccentricity distribution $\Delta n(E,e)$ for the stellar halo} \label{resultsDeltanEe}

The central part of deriving $\Delta n(E,e)$ lies in the observable time $\Delta T_r$. For the isochrone potential, after tedious algebra, we obtain the analytic expression of $\Delta T_r$ as follows:   
\begin{multline} \label{DTr_isochrone}
	\Delta T_r 
	= 2 \sqrt{\frac{b^3}{GM}} \theta 
	\left( \frac{\min(r_{\rm apo}, r_{+})}{b} - \frac{\max(r_{\rm peri}, r_{-})}{b} \right) \\
	\times 
	\left[ \frac{{\left( \varepsilon x^2 + 2 \sqrt{1+x^2} + \frac{{(1-e^2)}^2 - (1+e^2) \sqrt{ {(1-e^2)}^2 + 4\varepsilon^2 e^2}}{2\varepsilon e^2} 
	\right)}^{1/2}}{\varepsilon} 
		\right.\\
	\left. 
	+ \frac{\sin^{-1} \left( 
	\frac{ 2\sqrt{1+x^2} - \sqrt{1+x_{\rm apo}^2} - \sqrt{1+x_{\rm peri}^2 } }
	{\sqrt{1+x_{\rm apo}^2} - \sqrt{1+x_{\rm peri}^2} } 
	\right)}{{(-\varepsilon)}^{3/2}} 
	 \right]_{x=\max(r_{\rm peri}, r_{-})/b}^{x=\min(r_{\rm apo}, r_{+})/b} , 
\end{multline} 
where $\varepsilon \equiv 2bE/(GM)$, $x_{\rm peri} \equiv r_{\rm peri}/b$, and $x_{\rm apo} \equiv r_{\rm apo}/b$.\footnote{
Note that when $r_{-} < r_{\rm peri} < r_{\rm apo} < r_{+}$ is satisfied, the first term in the square bracket of equation (\ref{DTr_isochrone}) vanishes and the second term is equal to $\pi / {(-\varepsilon)}^{3/2}$, so that it gives $\Delta T_r = T_r = 2 \pi GM / {(-2E)}^{3/2}$ [cf. equation (33) of HY]. 
} 
Analytic expressions of $L^2$ and ${(\partial L^2 / \partial e)}_{E}$ 
in terms of $\varepsilon$ and $e$, which are necessary to derive $\Delta n(E,e)$, are found in equations (34) and (35) of HY. By substituting these quantities into equation (\ref{DeltanEe}) and using the parameters given in section \ref{pot}, we obtain the fully analytic expression of $\Delta n(E,e)$. 

Alternatively, for the NFW potential, we obtain the formal expression of $\Delta T_r$ as follows: 
\begin{multline}\label{DTr_NFW}
\Delta T_r = \sqrt{\frac{1}{2\pi G \rho_0}} 
\theta \left( \min(r_+, r_{\rm apo}) - \max(r_-, r_{\rm peri}) \right) \\
\int_{\max(r_-,r_{\rm peri})/a}^{\min(r_+,r_{\rm apo})/a} 
\frac{x dx}{\sqrt{\varepsilon x^2 + x\ln(1+x) -\frac{\lambda}{2}}}, 
\end{multline}
where $\varepsilon \equiv E/(4\pi G \rho_0 a^2)$ and  $\lambda \equiv L^2/(4\pi G \rho_0 a^4)$. Analytic expressions of other necessary quantities are found in equations (49) and (50) of HY, and we numerically calculate $\Delta n(E,e)$. 

\begin{figure*}
	\begin{center}
	\includegraphics[width=\columnwidth]{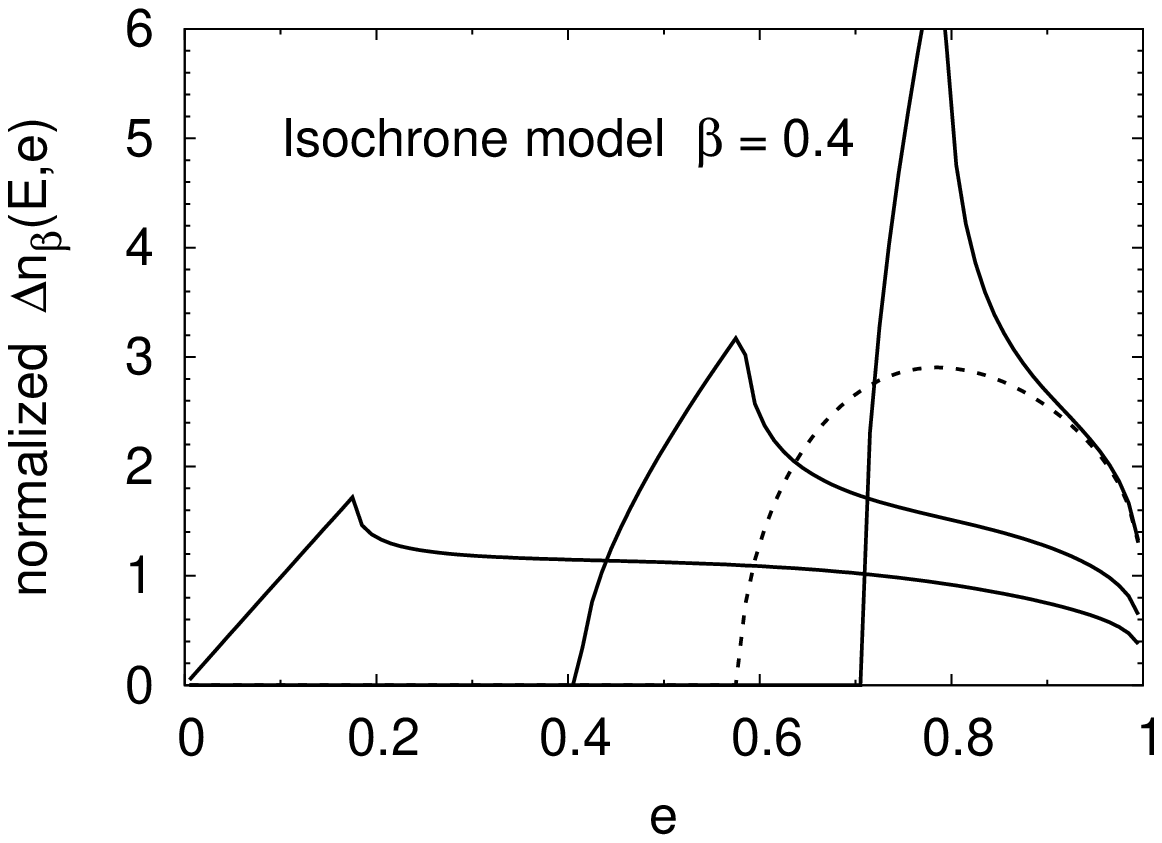}
	\includegraphics[width=\columnwidth]{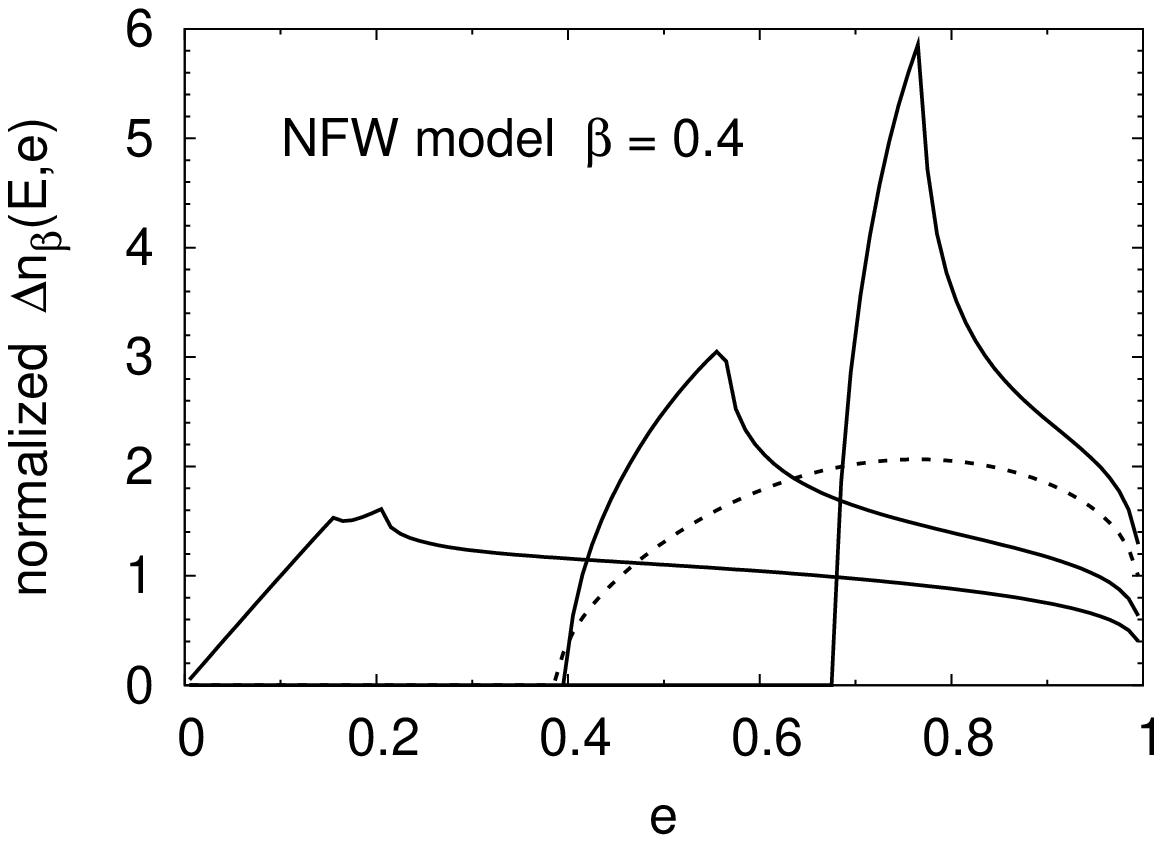}\\
	\includegraphics[width=\columnwidth]{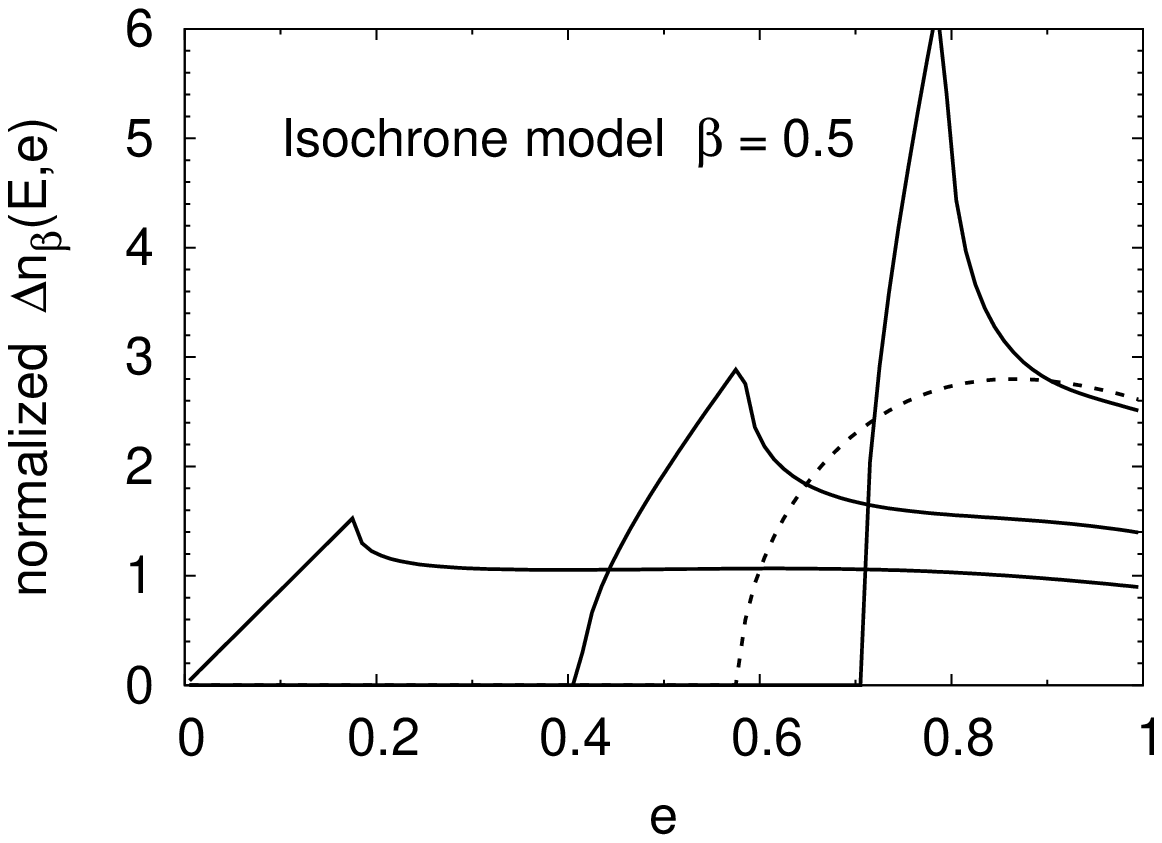}
	\includegraphics[width=\columnwidth]{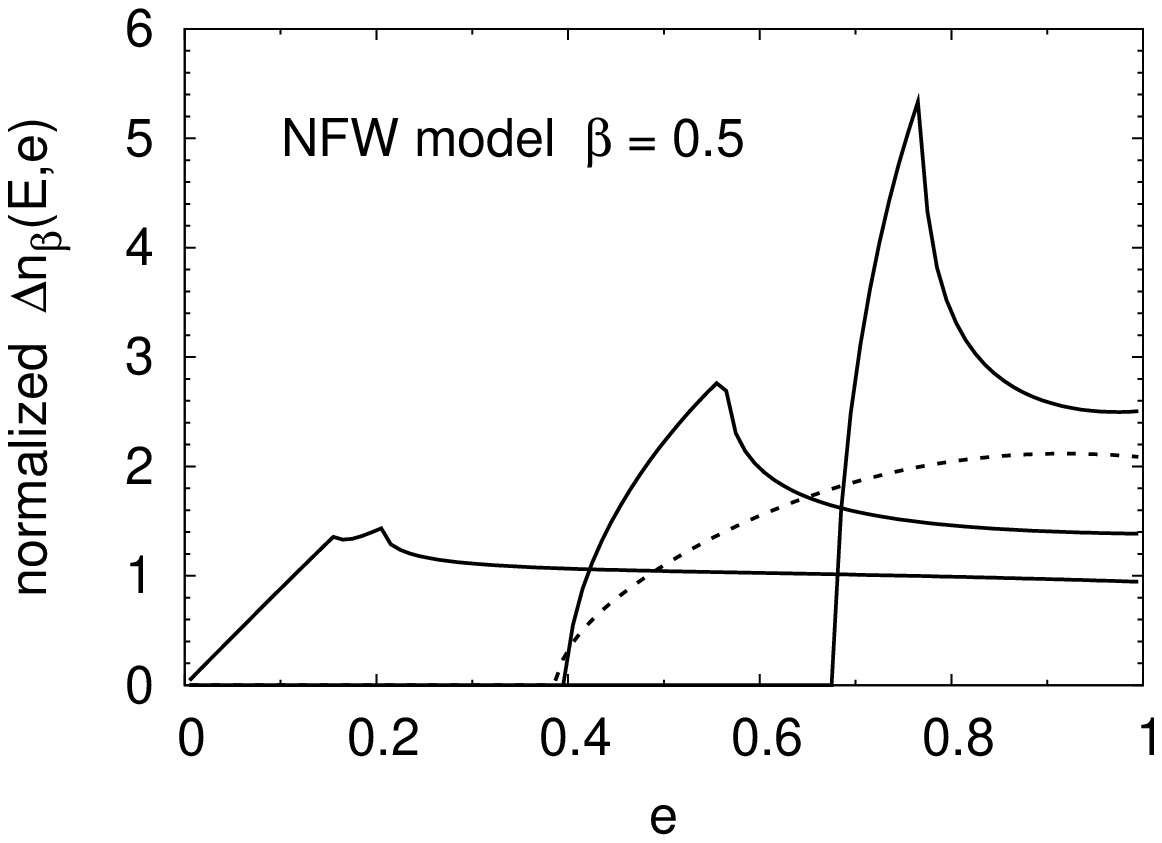}\\
	\includegraphics[width=\columnwidth]{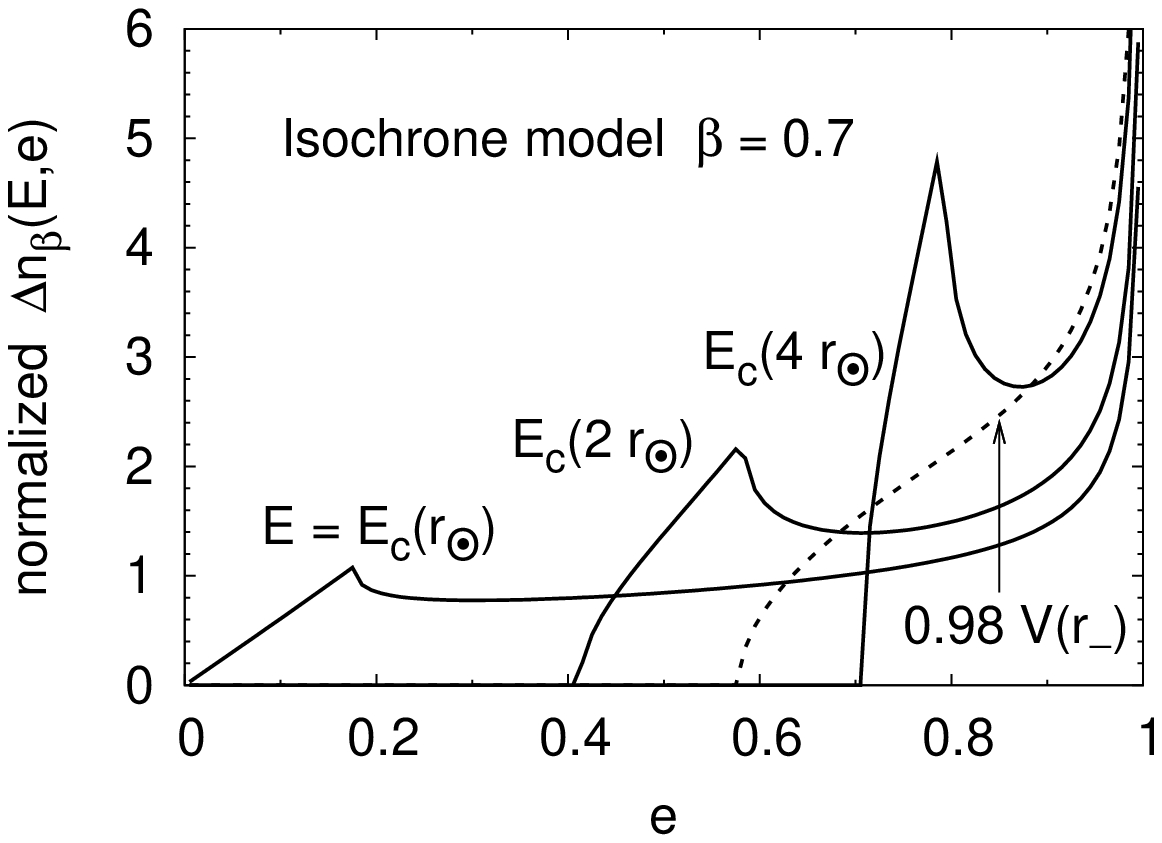}
	\includegraphics[width=\columnwidth]{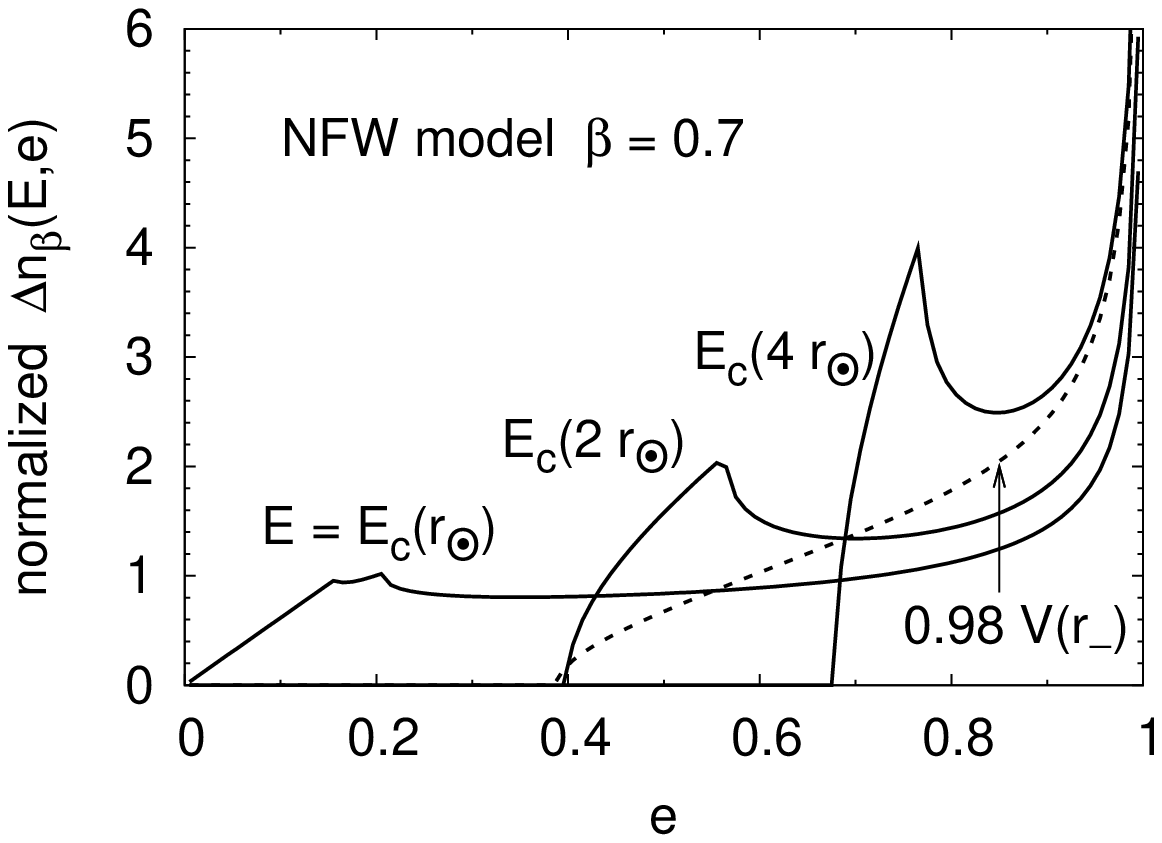}\\
	\caption{
Energy-dependent differential distribution $\Delta n_{\beta}(E,e)$ of stellar orbital eccentricity 
in a spherical system with constant radial $\beta$-profile for an observer 
whose survey region is $7\;{\rm kpc} < r <10\;{\rm kpc}$. 
Shown are the results in two realistic models of the gravitational potential for the Milky Way halo, 
such as the isochrone model on the left-hand panels and the NFW model in the right-hand panels. 
Shown by lines on each panel for given $\beta$ are the results for several values of specific energy $E$ 
equated to $E_c(r)$ of a star in circular orbit at different orbital radius $r$. 
Note that $\Delta n_{\beta}(E,e)$ is normalized such that $\int_0^1 \Delta n_{\beta}(E,e) de = 1$. 
}
\label{DeltanEeisoNFW}
	\end{center}
\end{figure*}

Figure \ref{DeltanEeisoNFW} shows calculations of $\Delta n_{\beta}(E,e)$ for constant profiles of $\beta=0.4, 0.5$, and $0.7$. 
Three panels of different values of $\beta$ in the left-hand column show the results for the isochrone potential, 
and three panels in the right-hand column for the NFW potential. 
In different panels for different values of $\beta$, shown by lines are the results for 
$E=0.98 V(r_{-}), E_{\rm c}(r_{\odot}), E_{\rm c}(2r_{\odot}),$ and $E_{\rm c}(4r_{\odot})$, 
where $E_{\rm c}(r)$ is defined in equation (\ref{Ec}). 
Irrespective of the potential adopted, we see common features in $\Delta n_{\beta}(E,e)$, 
i.e., its shape is weakly dependent on $\beta$ in the low-$e$ region, 
while the fraction of stars with high eccentricity notably increases as $\beta$ increases. 
We see from Figure \ref{DeltanEeisoNFW} that, just as in the case of thin-shell limit (Figure \ref{DeltanEeKepIHO}), 
$\Delta n_{\beta}$ has a non-zero cutoff eccentricity $e_{\rm cut}$ below which $\Delta n_{\beta}$ vanishes, 
when $E<E_{\rm c}(r_{-})$ and $E>E_{\rm c}(r_{+})$. 
This cutoff corresponds to 
$r_{\rm apo}=r_{-}$ when $E<E_{\rm c}(r_{-})$, and 
$r_{\rm peri}=r_{+}$ when $E>E_{\rm c}(r_{+})$. 
On the other hand, 
we also see that there are some eccentricities at which $\Delta n_{\beta}$ shows non-smooth behaviour. 
This singularity of $\Delta n_{\beta}$ corresponds to $r_{\rm apo} = r_{+}$ or $r_{\rm peri} = r_{-}$ 
and is analogous to the divergence of $\Delta n_{\beta}$ at $e=e_{\rm cut}$ in the thin-shell limit 
that corresponds to $r_{\odot}= r_{\rm peri}$ or $r_{\rm apo}$. 
The singularity associated with $r_{\rm apo}=r_{+}$ appears when $V(r_{+})<E<E_{\rm c}(r_{+})$ 
[e.g., $E=E_{\rm c}({r_{\odot}})$ in our results], while 
that associated with $r_{\rm peri}=r_{-}$ appears when $E_{\rm c}({r_{-}})<E$ 
[e.g., $E=E_{\rm c}({r_{\odot}})$, $E_{\rm c}(2{r_{\odot}})$, and $E_{\rm c}(4{r_{\odot}})$]. 
Both of these singularities appear in the overlapping $E$-region of 
$\max\{E_{\rm c}(r_{-}), V(r_{+})\} <E<E_{\rm c}(r_{+})$ [e.g., $E=E_{\rm c}({r_{\odot}})$]\footnote{
$\Delta n_{\beta}(E_c(r_{\odot}),e)$ has two singular points at $e\simeq0.2$ for both potential models. 
However, for the case of isochrone model, 
occasional proximity of these two singular points 
makes it difficult to see them separately in Figure \ref{DeltanEeisoNFW}.}, 
while none of them appears when $E<\min\{E_{\rm c}(r_{-}), V(r_{+})\}$ [e.g., $E=0.98 V(r_{-})$].

% iso+NFW bias   

\subsubsection{Eccentricity distribution $\Delta N(e)$ for the stellar halo} \label{resultsDeltaNe} 

Integrating $\Delta n(E,e)$ over $E$ [equation (\ref{DeltaNe})], we obtain $\Delta N(e)$ for the isochrone potential and the NFW potential. The results of $\Delta N(e)$ for both potentials similarly show a notable dependence on the profile of $\beta$. In the following, we summarize the characteristics of $\Delta N(e)$, which are common in both potentials except for some minor differences. 

\begin{figure*}
	\begin{center}
	\includegraphics[width=\columnwidth]{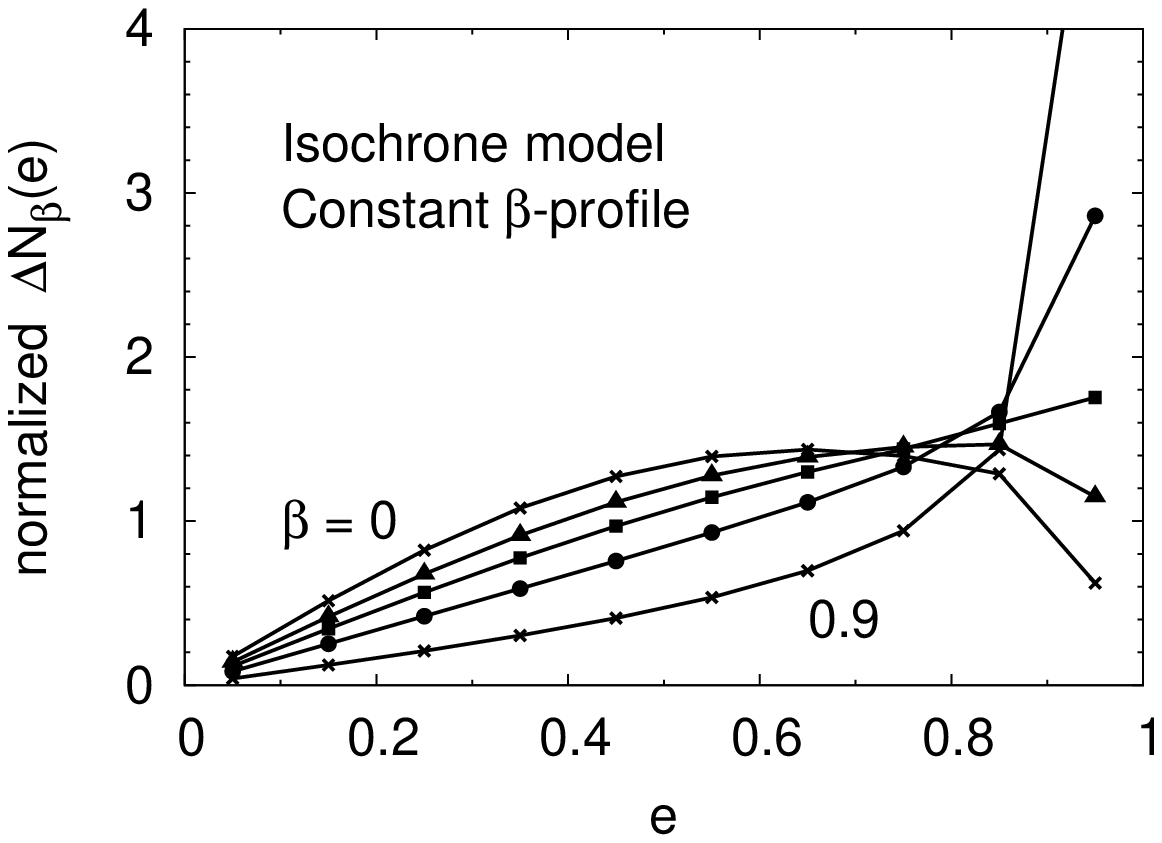}
	\includegraphics[width=\columnwidth]{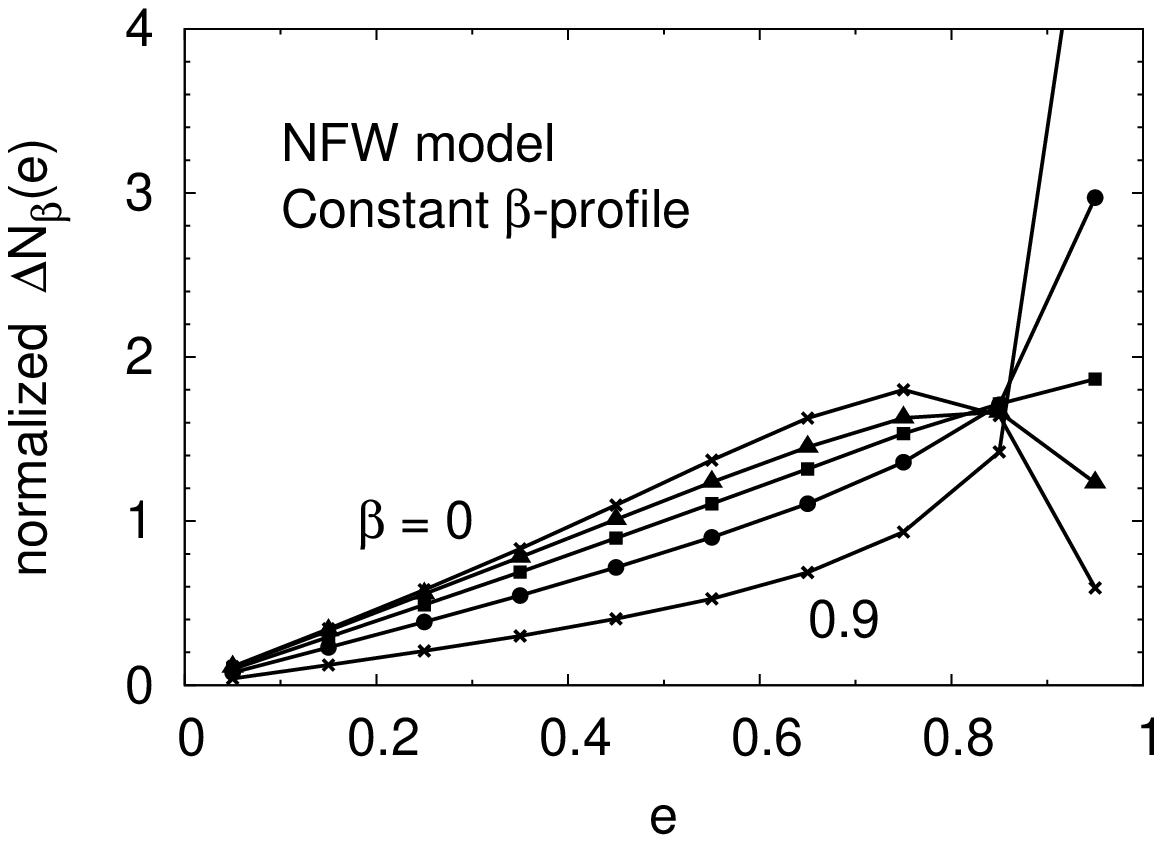}\\
	\caption{
Differential distribution $\Delta N_{\beta}(e)$ of stellar orbital eccentricity in a spherical system with constant radial $\beta$-profile for an observer whose survey region is $7\;{\rm kpc} < r <10\;{\rm kpc}$, embedded in the isochrone potential ({\it left-hand panel}) and the NFW potential ({\it right-hand panel}). 
Shown by lines on each panel are the results for $\beta=0$, 0.4, 0.55, 0.7, and 0.9, in order from top line to bottom line in the smaller-$e$ region. Note that $\Delta N_{\beta}(e)$ is normalized such that $\int_0^1 \Delta N_{\beta}(e) de = 1$. 
}\label{isoNFW_bias_sum_cst}
	\end{center}
\end{figure*}

% Fig.3
First, Figure \ref{isoNFW_bias_sum_cst} shows $\Delta N_{\beta}(e)$ with constant profile of $\beta$ for several values of $\beta$. A strong dependence of $\Delta N_{\beta}(e)$ on $\beta$ is apparent; As $\beta$ increases, $\Delta N_{\beta}(e)$ changes from a hump-like distribution ($\beta \lesssim 0.4$) through a nearly linear distribution ($\beta\sim 0.5-0.6$), then to a steeper turnup distribution with a prominent increase in the high-$e$ region ($\beta \gtrsim 0.7$). Particularly, as long as $\beta=0.4-0.7$, $\Delta N(e)$ is nearly linear in $e$ in the low-$e$ region ($e<0.5$). 

\begin{figure*}
	\begin{center}
	\includegraphics[width=\columnwidth]{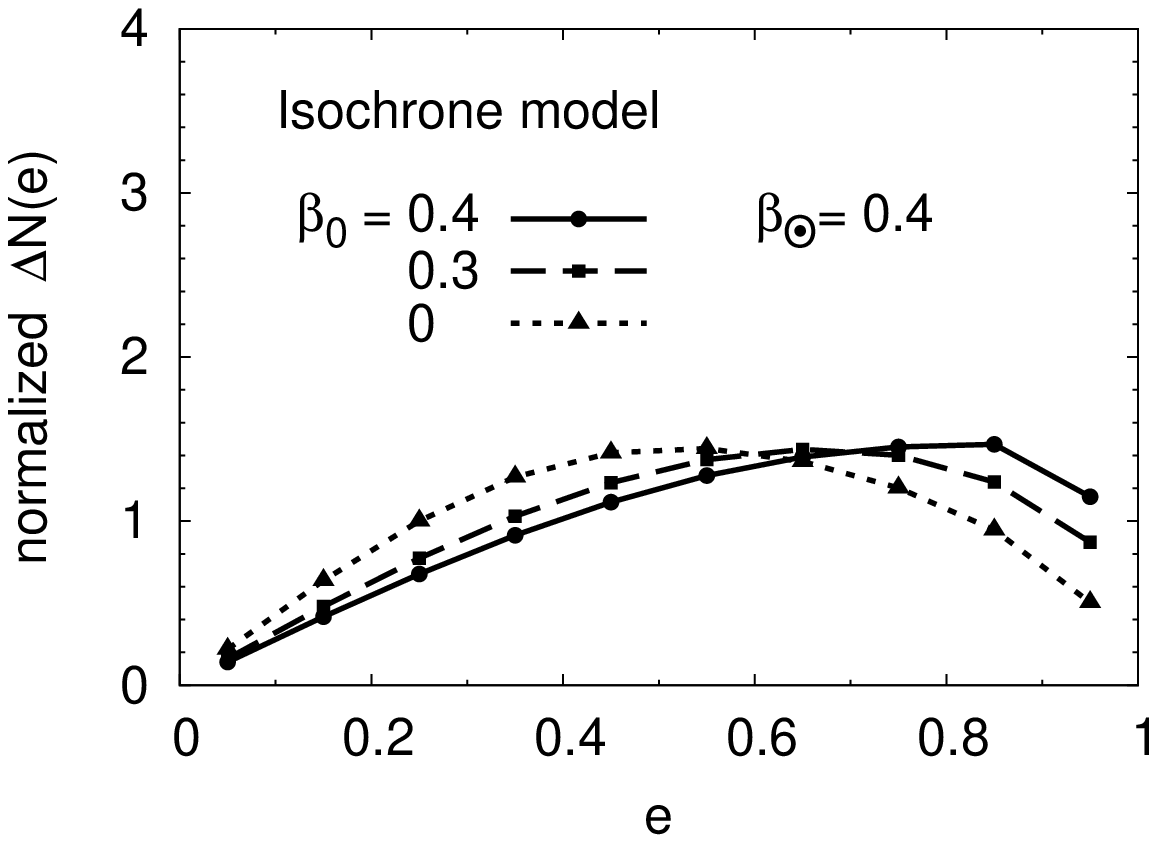}
	\includegraphics[width=\columnwidth]{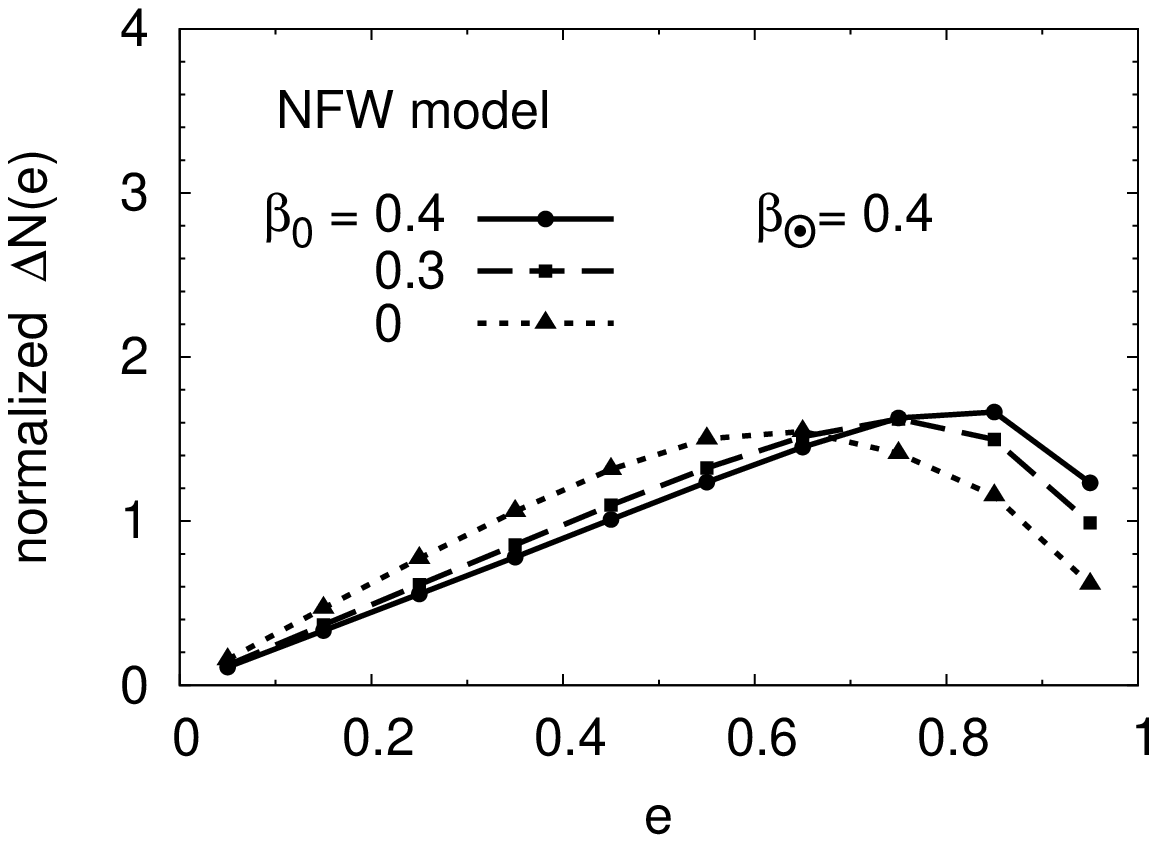}\\
	\includegraphics[width=\columnwidth]{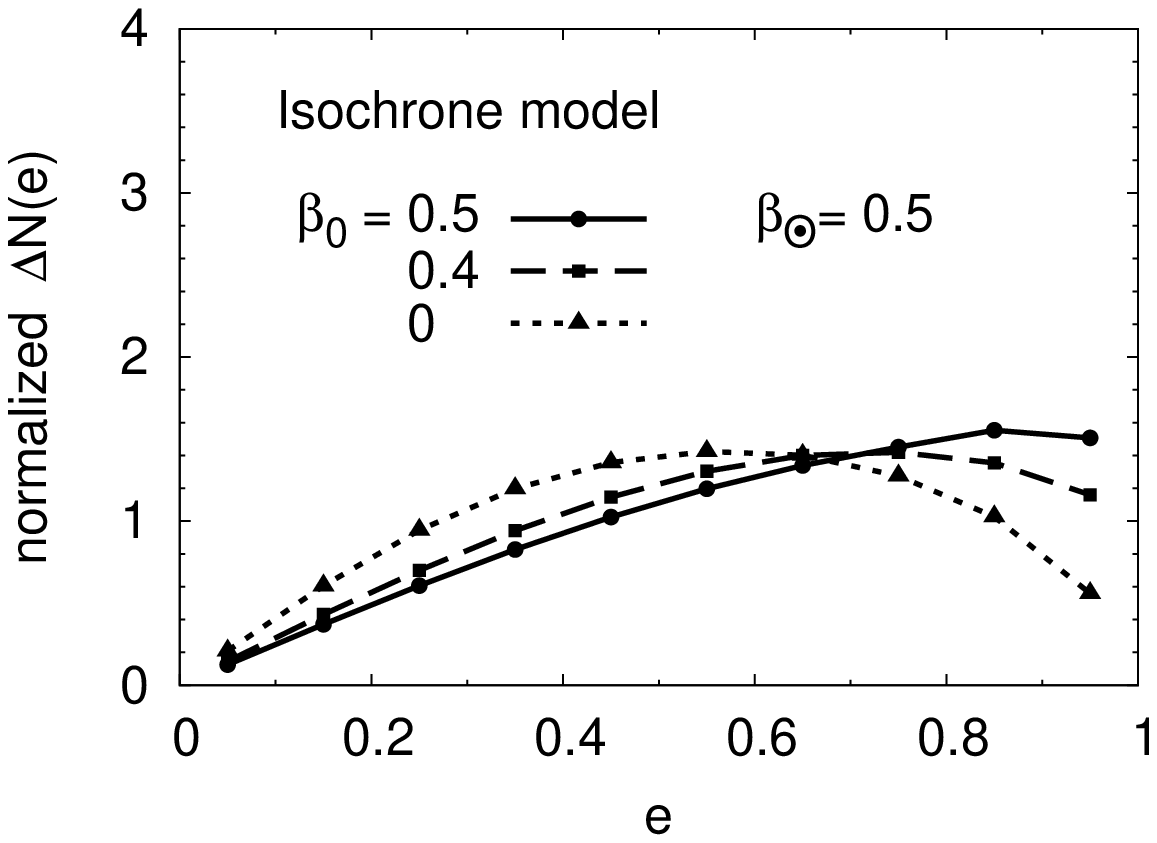}
	\includegraphics[width=\columnwidth]{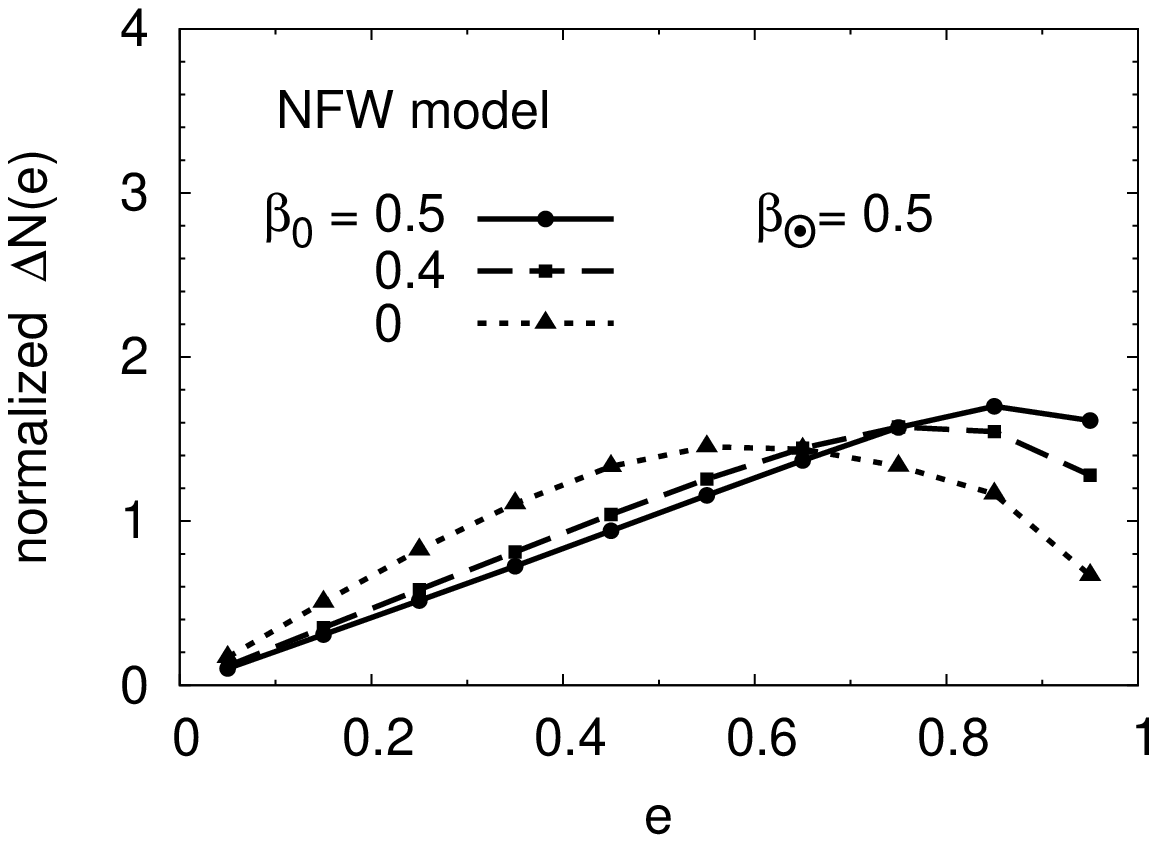}\\
	\includegraphics[width=\columnwidth]{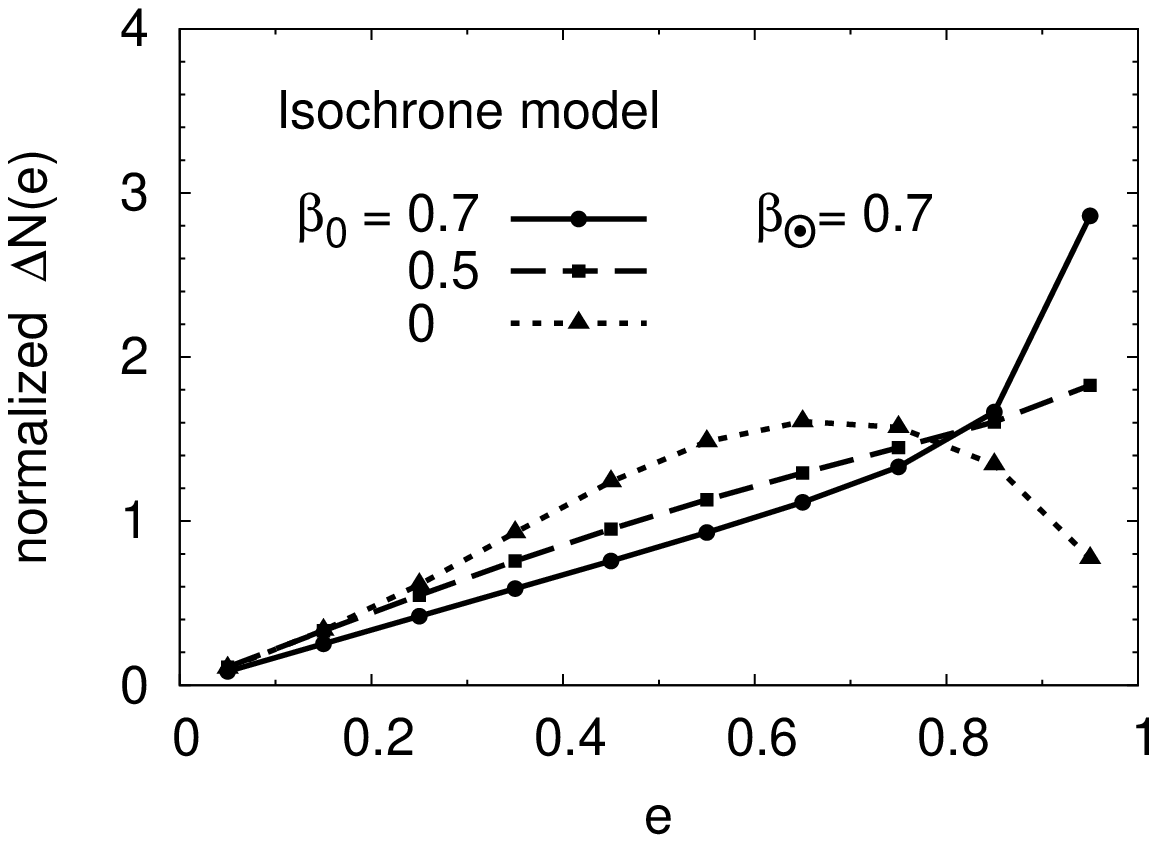}
	\includegraphics[width=\columnwidth]{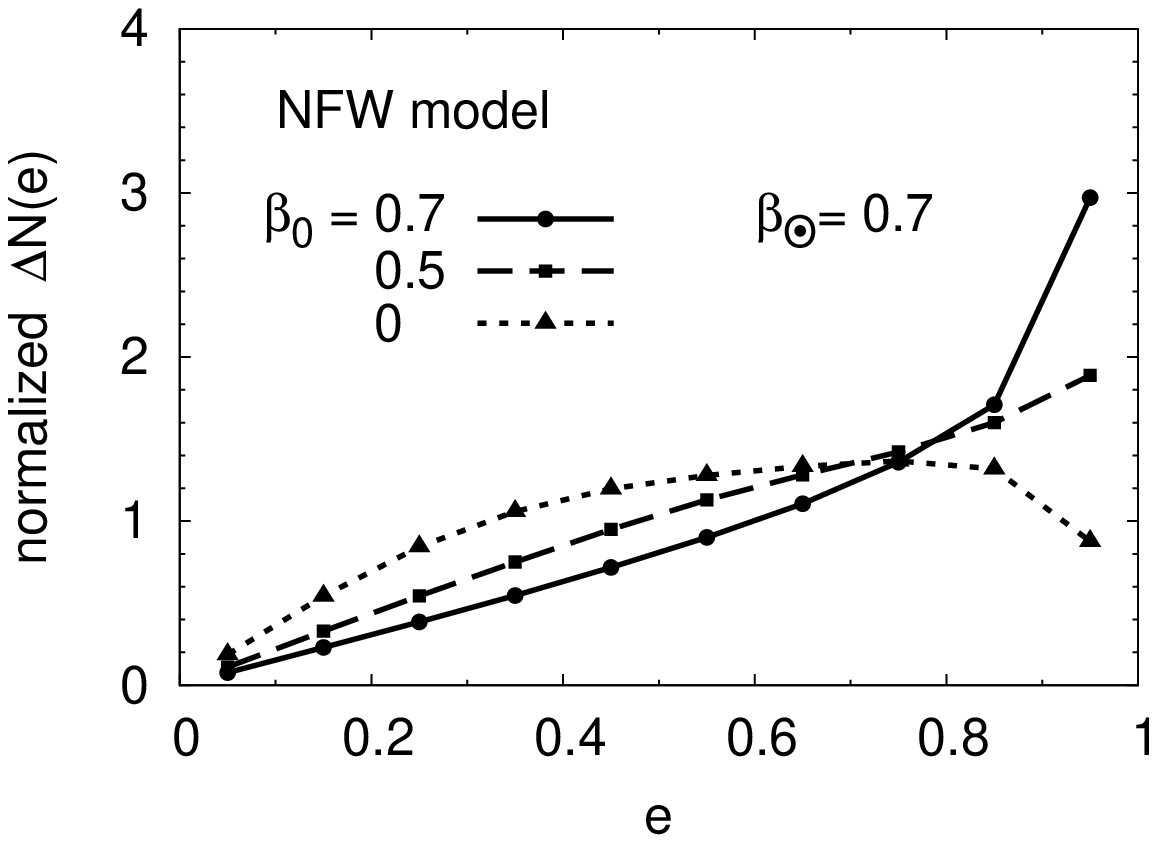}\\
	\caption{
Differential distribution $\Delta N_{\beta}(e)$ of stellar orbital eccentricity in a spherical system with the $\beta$-profile expressed by equation (\ref{betar}) for an observer whose survey region is $7\;{\rm kpc} < r <10\;{\rm kpc}$, embedded in the isochrone potential ({\it left-hand panel}) and the NFW potential ({\it right-hand panel}). 
Shown by dot-connecting lines on each panel for given $\beta_{\odot}$ are the results for several values of $\beta_0$. The case of $\beta_0 = \beta_{\odot}$ corresponds to the constant profile of $\beta$. In particular, the case of $\beta_0=0.5$ and $\beta_{\odot} = 0.7$ gives the nearly linear $e$-distribution over a full range of $e$ for both potentials adopted. Note that $\Delta N(e)$ is normalized such that $\int_0^1 \Delta N(e) de = 1$. 
}\label{isoNFW_bias_sum}
	\end{center}
\end{figure*}

% Fig.4
Second, Figure \ref{isoNFW_bias_sum} shows $\Delta N(e)$ with non-constant profile of $\beta$ for several combinations of ($\beta_0, \beta_\odot$). The linear trend of $\Delta N(e)$ in the low-$e$ region, which is seen in Figure \ref{isoNFW_bias_sum_cst}, remains as long as $\beta_{\odot}=0.4-0.7$ and $\beta_0 \gtrsim 0$. Figure \ref{isoNFW_bias_sum} also shows that the fraction of high-$e$ stars increases as $\beta_0$ increases from $0$ to $\beta_{\odot}$. In particular, the fraction of stars with $0.9<e<1$ for $\beta_0=\beta_{\odot}$ is nearly 2-3 times larger than that for $\beta_0=0$. 

\begin{figure*}
	\begin{center}
	\includegraphics[width=\columnwidth]{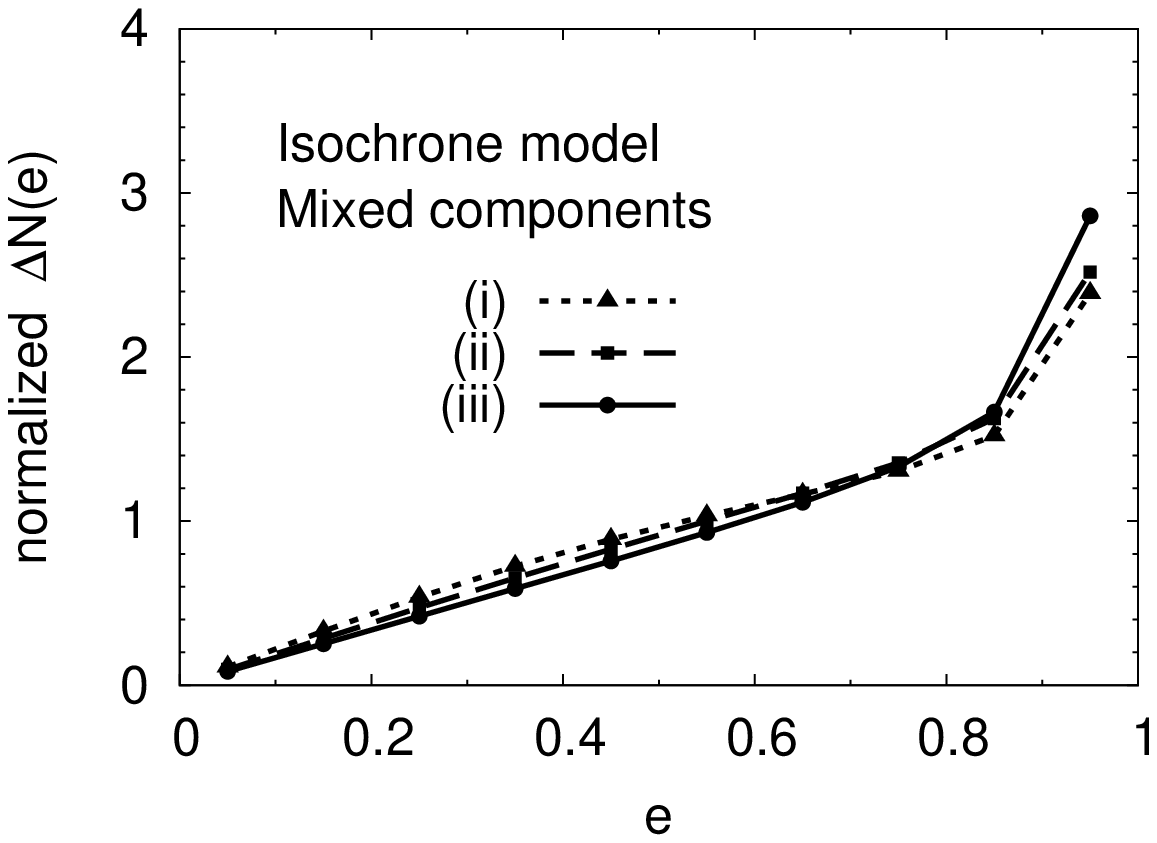}
	\includegraphics[width=\columnwidth]{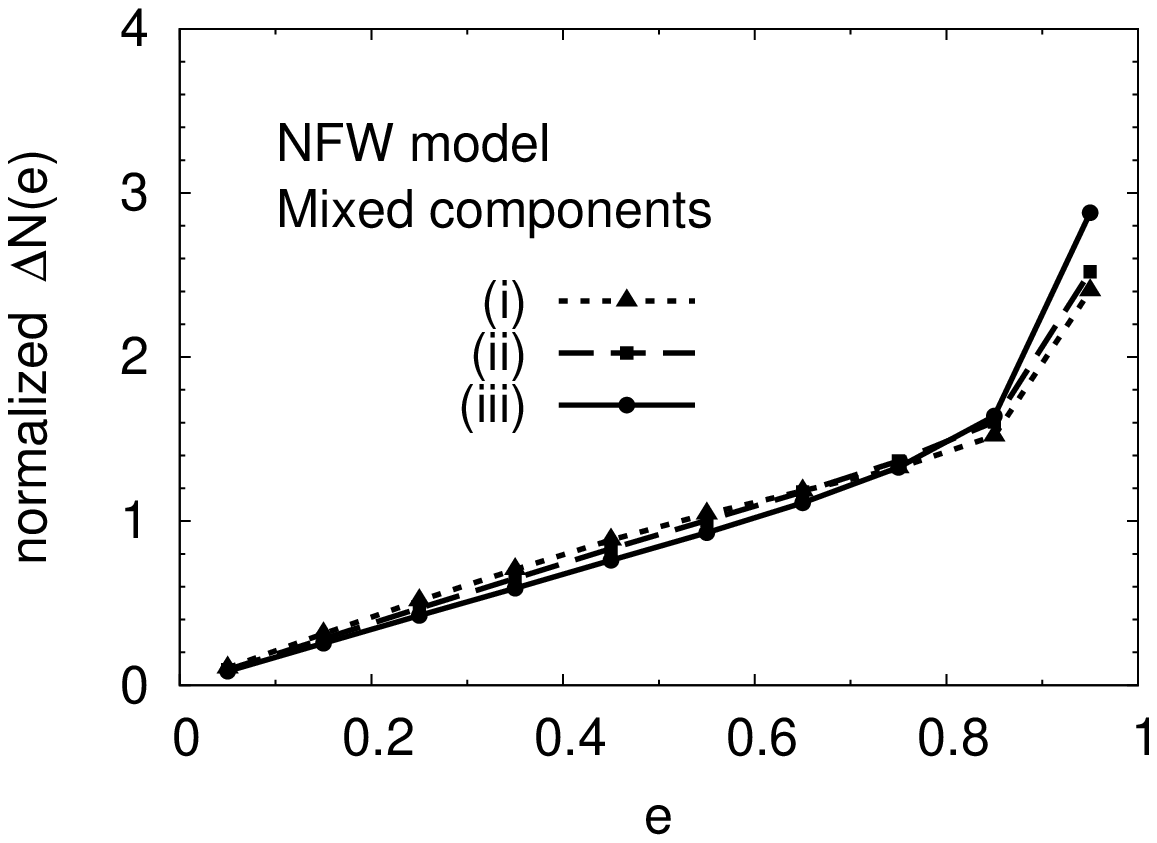}\\
	\caption{
Differential distribution $\Delta N_{\beta}(e)$ of stellar orbital eccentricity for stars in $7\;{\rm kpc} < r <10\;{\rm kpc}$ that are made up of two stellar components, embedded in the isochrone potential ({\it left-hand panel}) and the NFW potential ({\it right-hand panel}). 
Shown are the results for three cases such as 
(i) a mixture of an 80\% component with ($\beta_0,\beta_\odot$)=(0.7,0.7) and 
a 20\% component with (0.0,0.4), 
(ii) a mixture of an 80\% component with (0.7,0.7) and 
a 20\% component with (0.4,0.4), and (iii) a single component with (0.7,0.7). Note that $\Delta N(e)$ is normalized such that $\int_0^1 \Delta N(e) de = 1$. 
}\label{isoNFW_bias_sum_two}
	\end{center}
\end{figure*}

% Fig.5
Third, Figure \ref{isoNFW_bias_sum_two} shows $\Delta N(e)$ made up of two stellar components. 
Following the recent observation by \cite{Carollo2010} that the Milky Way halo consists of at least two components, 
such as the inner halo with $\beta_{\odot}\simeq 0.7$ and the outer halo with $\beta_{\odot}\simeq 0.4$, 
we here consider three cases: 
(i) a mixture of an 80\% component with ($\beta_0,\beta_\odot$)=(0.7,0.7) and 
a 20\% component with (0.0,0.4), 
(ii) a mixture of an 80\% component with (0.7,0.7) and 
a 20\% component with (0.4,0.4), and 
(iii) a single component with (0.7,0.7). 
Calculating the eccentricity distributions for individual components, 
and adding them up according to the mixture ratio adopted, 
we obtain $\Delta N(e)$ for each of the cases (i) and (ii), to be compared with the case (iii). 
We see from this figure that contamination of the lower-$\beta_\odot$ outer-halo component by 20\% level 
gives no significant effect on $\Delta N(e)$ for the higher-$\beta_\odot$ inner-halo component, particularly, at $e \gtrsim 0.7$. 
In other words, from the observed $\Delta N(e)$ we can best constrain the profile of $\beta$ for the inner-halo component, 
in a manner largely unaffected by possible contamination of the outer-halo component.

% isochrone+NFW bias beta=const

% isochrone+NFW bias  

% isochrone+NFW bias two-components 

\section{Discussion} 

In the previous section, we have performed theoretical calculations of $\Delta N(e)$ for the stellar halo, 
assuming spherical symmetry for simplicity. 
In this section, we demonstrate how our calculations would be useful 
in interpreting the observations of kinematics of halo stars in the Milky Way. 

\subsection{Comparison with the observed eccentricity distribution}
\subsubsection{Summary of the latest observation} \label{latest} 
So far the largest, kinematically unbiased sample of halo stars is that obtained by \cite{Carollo2010}. They evaluated eccentricities of solar-neighbour halo stars sampled in $7\;{\rm kpc}<R<10\;{\rm kpc}$ and $d<4\;{\rm kpc}$, 
using an axisymmetric St${\rm \ddot a}$ckel-type potential 
which was first used by \cite{SLZ1990} for the analysis of stellar halo. 
Their finding is that the eccentricity distribution of stars differs significantly 
depending on the metallicity range chosen (see their Figure 5). 
The results for likely-halo stars with [Fe/H]$<-1.5$ and $|z|>1\;{\rm kpc}$ 
($|z|$ is the distance from the disk plane) are summarized as follows: 
(1) $\Delta N_{\rm obs}(e)$ for stars with $-2.0<$[Fe/H]$<-1.5$ shows a linear trend over the full range of eccentricity from $e=0$ to 1. 
(2) $\Delta N_{\rm obs}(e)$ for stars with [Fe/H]$<-2.0$ also shows a linear trend at $e<0.6-0.8$, but with a shallower slope toward higher $e$, in contrast with the case of $-2.0<$[Fe/H]$<-1.5$. 
(3) $\Delta N_{\rm obs}(e)$ as above is unaffected by the spatial criterion for sampling stars at either $1\;{\rm kpc}<|z|<2\;{\rm kpc}$ or $2\;{\rm kpc}<|z|<4\;{\rm kpc}$.

\cite{Carollo2010} expected that this metallicity-dependent $\Delta N_{\rm obs}(e)$ 
could be explained by the metallicity dependence of mixture ratio of the inner- and the outer-halo components. According to their decomposition analysis, the sample for $-2.0<$[Fe/H]$<-1.5$ is mainly ($\gtrsim 80$\%) contributed by the inner-halo component. Although they did not perform a similar decomposition analysis on the sample with [Fe/H]$<-2.0$, they supposed a smaller fraction of high-$e$ stars in this very metal-poor sample than that for $-2.0<$[Fe/H]$<-1.5$, because of larger contamination of outer-halo component which is expected to possess a smaller fraction of high-$e$ stars than the inner-halo component. 

By comparing all these latest results of $\Delta N_{\rm obs}(e)$ with our theoretical calculations of $\Delta N(e)$, we discuss their implications in the global kinematical structure below.

\subsubsection{Linear trend in the low-eccentricity region}
A linear $e$-distribution obtained by \cite{Carollo2010} prevails at least at $e<0.6$, which is common to both samples of $-2<$[Fe/H]$<-1.5$ and [Fe/H]$<-2.0$ [results (1) and (2) in section \ref{latest}]. Combined with their estimation of $\beta_{\odot}\simeq 0.7$ for the inner halo and $\beta_{\odot}\simeq0.4$ for the outer halo,\footnote{
These values of $\beta_{\odot}$ are derived from $(\sigma_r, \sigma_{\theta}, \sigma_{\phi}) = (160,102,83)$ ${\rm km\;s^{-1}}$ for the inner halo, and $(178,149,127)$ ${\rm km\;s^{-1}}$ for the outer halo, which were reported in \cite{Carollo2010}.
} 
the observed linear trend in the low-$e$ region is well reproduced by our theoretical $e$-distribution $\Delta N(e)$ for both of the isochrone and NFW potentials, provided  $\beta_0 \gtrsim 0$. We expect that this result holds for any realistic {\it spherical} mass distribution models of the Milky Way halo, because the isochrone and NFW potentials form two extreme ends of such models. In addition, it may as well hold for {\it non-spherical} mass distribution models, because the equi-potential surface is rounder than the equi-density surface and the estimation of eccentricity is hardly affected by the non-sphericity of the potential at least in the low-$e$ region. 
Thus, regardless of the mixture ratio of the inner halo ($\beta_{\odot}\simeq 0.7$) and outer halo ($\beta_{\odot}\simeq 0.4$) and regardless of either the spherical or non-spherical mass distribution assumed for the halo, we still expect that a nearly linear $e$-distribution would occur at $e<0.5-0.6$.

\subsubsection{A constraint on the radial $\beta$-profile for the inner halo} \label{IHconstraint}
Figure 5 of \cite{Carollo2010} shows that the linear trend of $\Delta N_{\rm obs}(e)$ extends beyond the low-$e$ region up to $e=1$ for a sample of stars with $-2.0<$[Fe/H]$<-1.5$ [result (1) in section \ref{latest}]. This trend can be used to constrain the radial profile of $\beta$ for the inner halo which dominantly contributes to such a sample. 

In general, when a non-spherical potential is adopted, just as in \cite{Carollo2010}, the estimation of eccentricity of a star tends to be systematically larger than that for a spherical potential. This discrepancy cannot be ignored at high-$e$ region, so that adopting a non-spherical potential instead of a spherical one would enhance the fraction of high-$e$ stars, while suppressing the fraction of low-$e$ stars. Accordingly, the approximately linear shape of $\Delta N_{\rm obs}(e)$ up to $e=1$ for a sample of likely inner-halo stars with $-2.0<$[Fe/H]$<-1.5$, reported by \cite{Carollo2010}, would be changed into a {\it hump-like} shape when the eccentricity in their analysis is instead defined in our spherical model of the Milky Way halo. We then constrain the radial profile of $\beta$ for the inner halo by examining how our spherical model could reproduce such a hump-like shape of $\Delta N_{\rm obs}(e)$.  

Noting $\beta_{\odot}\simeq0.7$ for the inner halo, our models of $\Delta N(e)$ for various $\beta$-profiles with $\beta_{\odot}=0.7$, shown on the bottom-row panels of Figure \ref{isoNFW_bias_sum}, would help constrain the $\beta$-profile for the inner halo. We see from these panels that our models for more or less constant $\beta$-profiles with $\beta_0=0.5-0.7$ and $\beta_{\odot}=0.7$ would be ruled out. On the other hand, our models for notably $r$-dependent $\beta$-profiles with $\beta_0\lesssim 0.5$ and $\beta_{\odot}=0.7$ would reproduce $\Delta N_{\rm obs}(e)$ if a {\it non-spherical} potential is assumed to be consistent with Carollo et al.'s analysis. Remembering $(r_a/r_{\odot})^2=(1-\beta_{\odot})/(\beta_{\odot}-\beta_0)$, a constraint of $\beta_0 \lesssim 0.5$ for the inner halo is equivalent to $r_a \lesssim 10\;{\rm kpc}$. 

The key idea used above is that the sample of stars with $-2.0<$[Fe/H]$<-1.5$ by \cite{Carollo2010} has a smaller fraction of high-$e$ stars when compared with that expected from our models with a constant profile of $\beta_{\odot}=0.7$. Since smaller $\beta_{\odot}$ results in a smaller fraction of high-$e$ stars (see Figure \ref{isoNFW_bias_sum}), some contamination of a stellar component with smaller $\beta_{\odot}$ would obviously decrease the fraction of high-$e$ stars in the sample. Thus, we examine whether the observed small fraction of high-$e$ stars can be explained by our models if the contamination of the outer halo with $\beta_{\odot}\simeq 0.4$ is taken into account. Figure \ref{isoNFW_bias_sum_two} clearly shows this effect. For the cases (i) and (ii) in section \ref{resultsDeltaNe} where 80\% of the sample is contributed by the inner halo with the constant profile of $\beta_{\odot}=0.7$ and the other 20\% by the outer halo with $\beta_{\odot}=0.4$, the fraction of high-$e$ stars in the sample becomes distinguishably smaller than that for the case (iii) for which the sample genuinely consists of the inner halo. However, we see from this figure that 20\% contamination of the outer halo is not able to make a hump-like shape of $\Delta N(e)$ and is not enough to explain the observed small fraction of high-$e$ stars in the sample. Therefore, the observed $e$-distribution reported by \cite{Carollo2010} implies that the radial $\beta$-profile for the inner halo is not constant. Rather, it is consistent with a $\beta$-profile that increases away from the galaxy center with $\beta_0 \lesssim 0.5$.

\subsection{Insights into the formation of the Milky Way}

In the previous subsection, we have demonstrated a non-constant radial $\beta$-profile in the inner-halo component. In this subsection, we briefly discuss the insights of this result into the formation of the Milky Way halo.

\subsubsection{Radial $\beta$-profile as a probe of the relaxation process}

\cite{Lynden-Bell1967} argued that a stellar system that experienced a rapid change of the gravitational potential would evolve into the relaxed system with an isotropic, ergodic distribution function. He called this phenomenon `violent relaxation.' He expected that if the relaxation process is spatially limited in the central part of the system, the resultant distribution function would be altered so that it is isotropic only at the center, while staying anisotropic in the outer part of the system. As an example that could approximately represent such a system, he mentioned a distribution function of the form $f(E,L)=f(Q)$, which was later shown to accommodate a radial profile of $\beta = r^2/(r^2+r_a^2)$ (\citealt{Osipkov1979}; \citealt{Merritt1985}). In his view, the scale radius $r_a$ corresponds to the `relaxation radius' inside which the relaxation process acts effectively, in reasonable agreement with the numerical experiments of violent relaxation (e.g., \citealt{vanAlbada1982}). 

The distribution function we used in this paper [equation (\ref{DFB})] is a natural generalization of $f(Q)$, and allows a free parameter of $\beta_0$ at the center of the system in addition to $r_a$. If the actual $\beta$-profile is described by the form in equation (\ref{betar}), estimates of $r_a$ or $\beta_0$ may be used as a guide to understand the relaxation process that has acted on the system. 
For example, if the scale radius $r_a$ could indicate the spatial reach of effective relaxation, as noted by \cite{Lynden-Bell1967}, our present constraint of $r_a \lesssim 10\;{\rm kpc}$ or equivalently $\beta_0 \lesssim 0.5$ for the inner halo (section \ref{IHconstraint}) might serve as a hint to uncover a trigger of such relaxation in the early Galaxy. Moreover, if some notable deviation of $\beta_0$ from zero could be seen, it might indicate that the relaxation would not be effective enough to completely erase the initial dynamical condition even at the center of the system. In order to impose a stronger constraint on $\beta_0$ than our present constraint of $\beta_0 \lesssim 0.5$ (section \ref{IHconstraint}), it is necessary to re-analyse a sample by \cite{Carollo2010} and derive $\Delta N_{\rm obs}(e)$ in a {\it spherical} halo potential, with which we can directly compare our models. 

We note that our present constraint on the large-scale profile of $\beta$ beyond the solar neighbourhood should be tested against direct measurements of $\beta$ at various radial distances away from the center. 
In fact, \cite{Bond2010} %Bond et al. (2010) %
reported that such analysis for a sample of stars with [Fe/H]$<-1.1$ in $3\;{\rm kpc}<R<13\;{\rm kpc}$ and $1\;{\rm kpc}<|z|<5\;{\rm kpc}$ supports a nearly constant profile of $\beta$. 
More works for direct measurements of $\beta$ are obviously necessary for further discussions on its radial profile in the inner halo. 
In the meanwhile, we plan to extend our present analysis to include another type of distribution functions 
in which $\beta_0 > \beta_{\odot}$ is also allowed (Hattori 2011, in preparation). 
With these calculations, we would be able to test whether $\beta(r)$ is a decreasing function of $r$ 
(as proposed by \citealt{SL1997}) or not. 
However, we stress that our present calculations of $\Delta N(e)$ 
do help constrain the large-scale profile of $\beta$ from a sample of solar-neighbour halo stars.

\section{Conclusion}

In this paper, we formulate the eccentricity distribution of the solar-neighbour halo stars $\Delta N(e)$, with simple assumption of the spherical halo. By adopting two appropriate halo potentials, we show that $\Delta N(e)$ is highly dependent on the radial profile of velocity anisotropy parameter $\beta$. Moreover, we show that our theoretical calculations of $\Delta N(e)$ are useful in explaining some properties of $\Delta N_{\rm obs}(e)$ by \cite{Carollo2010}, such as the linear $e$-distribution in the low-$e$ region and the metallicity-dependent fraction of high-$e$ stars.

We have demonstrated that the observed fraction of high-$e$ stars with $-2.0<$[Fe/H]$<-1.5$ by \cite{Carollo2010} is smaller than that expected from our models with a constant profile of $\beta$ in the inner halo. This result places a constraint on its radial profile for the inner-halo component of the Milky Way, yielding $\beta_0 \lesssim 0.5$ at the galaxy center in contrast to the obsered value of $\beta_{\odot}\simeq0.7$ in the solar neighbourhood. This result further shows that the scale radius $r_a$ of the $\beta$-profile should be smaller than $\sim 10\;{\rm kpc}$, which might imply that some relaxation process that acted on the inner-halo component was effective only within this radius away from the center of the Milky Way.

\section*{Acknowledgements}
We thank Beers, T., Carollo, D., Minezaki, T., Tsujimoto, T., Yamagata, T., Sakata, Y., Kakehata, T., and Fujii, H. for useful discussions and suggestions. 
KH is supported by JSPS Research Fellowship for Young Scientists (23$\cdot$954), and partly by Hayakawa Sachio Foundation.

%##########################################################################

%\bibliographystyle{mn}
\bibliography{mn-jour,refHY2011}

\appendix

\section{Lindblad diagram} \label{Lindblad} 
The aim of this appendix is to show the phase-space region in which observable stars are distributed and to demonstrate how the cutoff eccentricity $e_{\rm cut}$ depends on specific energy $E$.

Let us consider a spherical system of stars in a spherical gravitational potential $V(r)$, and an observer $O_1$ whose survey region is $r_{-}<r<r_{+}$.\footnote{
This assumption of survey region may seem too simple, but is justified by considering a more realistic observer $O_2$ at $r = (r_{+}+r_{-})/2$ who observes stars within a distance $({r_{+}-r_{-}})/2$ away from him. Since the survey region of observer $O_1$ includes that of observer $O_2$, the latter region well represents the former as long as $({r_{+}-r_{-}})/2$ is small. Accordingly, the eccentricity distribution for these observers would be similar to each other in this case.
\label{fn}}
As noted in section \ref{formulation}, observable stars for $O_1$ are defined as those satisfying $r_{-} < r_{\rm apo}$ and $r_{\rm peri}<r_{+}$. Since $r_{\rm apo}$ and $r_{\rm peri}$ depend on $E$ and $L$, the phase space region which is occupied by observable stars can be clearly shown in the ($E,L$)-phase space, called `Lindblad diagram' (\citealt{May1986}).

\begin{figure}
	\begin{center}
	\includegraphics[width=\columnwidth]{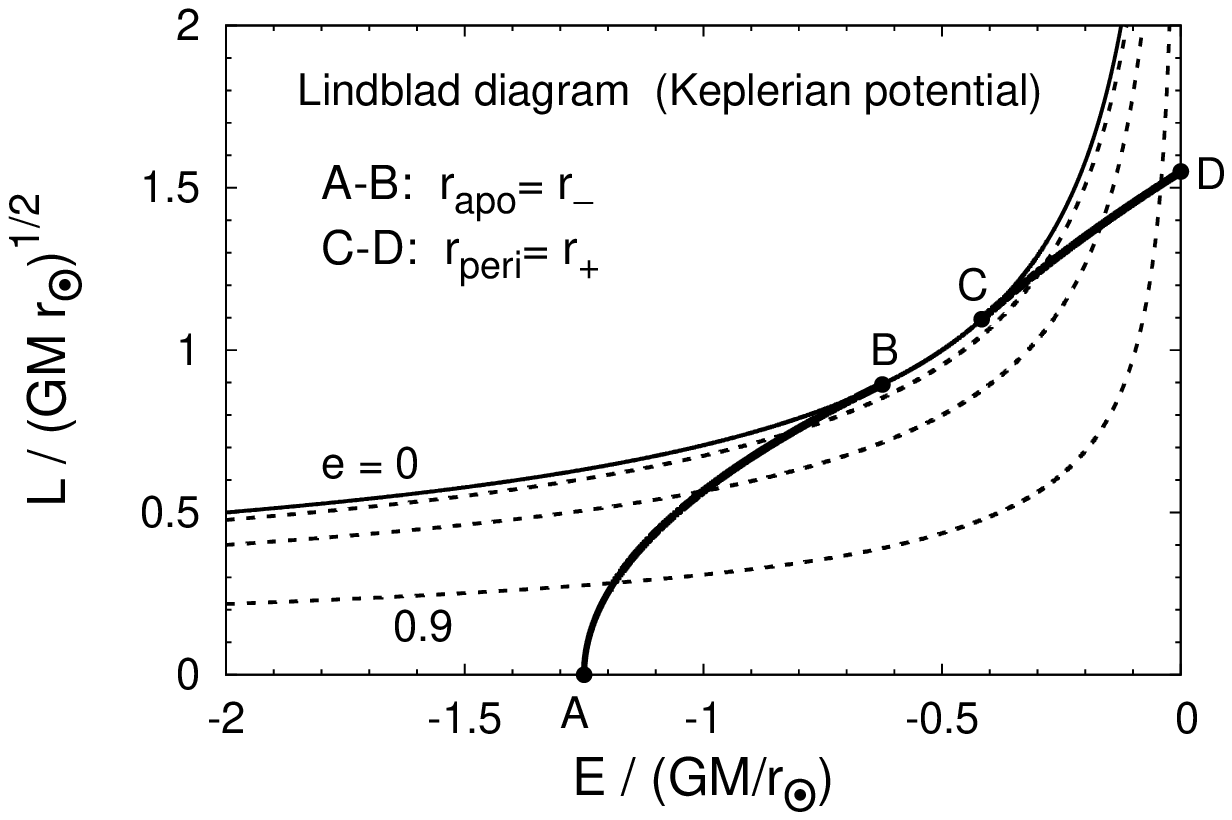}
        \caption{
Lindblad diagram with the Keplerian potential $V(r)=-GM/r$ for an observer at $r=r_{\odot}$ whose survey region is $r_{-}<r<r_{+}$. 
In this case, we assume $r_{-}=0.8 r_{\odot}$ and $r_{+}=1.2 r_{\odot}$. 
The abscissa represents the specific energy $E$ in units of $GM/r_{\odot}$ and 
the ordinate represents the specific angular momentum $L$ in units of $\sqrt{GMr_{\odot}}$. 
No stars are allowed above the contour of $e=0$ (solid thin line), 
while observable stars are distributed below A-B-C-D (solid thick line), 
where A-B represents the contour of $r_{\rm apo}=r_{-}$ and C-D represents the contour of $r_{\rm peri}=r_{+}$. 
Here, the positions of A, B, C, and D correspond to $E=V(r_{-})$, $E_{\rm c}(r_{-})$, $E_{\rm c}(r_{+})$, and 0, respectively. The contours of $e=0.3, 0.6,$ and $0.9$ are shown by dashed thin lines, 
which help understand that the cutoff eccentricity $e_{\rm cut}$ decreases with increasing $E$ from A to B [$V(r_{-})<E<E_{\rm c}(r_{-})$], remains zero from B to C [$E_{\rm c}(r_{-})<E<E_{\rm c}(r_{+})$], and increases from C to D [$E_{\rm c}(r_{+})<E<0$]. 
}\label{LindKep}
	\end{center}
\end{figure}

Figure \ref{LindKep} is an example of such diagram for the Keplerian potential.
We see from this diagram that stars with $E<V(r_{-})$ are unobservable ($\Delta T_r=0$) regardless of $e$ ($e_{\rm cut}=1$) because $r_{\rm apo}<r_{-}$, while stars with $E_{\rm c}(r_{-})<E<E_{\rm c}(r_{+})$ are observable ($\Delta T_r>0$) regardless of $e$ ($e_{\rm cut}=0$) because both $r_{-}<r_{\rm apo}$ and $r_{\rm peri}<r_{+}$ are satisfied. On the other hand, stars with $V(r_{-})<E<E_{\rm c}(r_{-})$ are observable 
only when $e>e_{\rm cut}$ where $e_{\rm cut}$ corresponds to $r_{\rm apo}=r_{-}$.
In this range of $E$, $e_{\rm cut}$ decreases monotonically from 1 at $E=V(r_{-})$ to 0 at $E=E_{\rm c}(r_{-})$. Similarly, stars with $E_{\rm c}(r_{+})<E<0$ are observable only when $e>e_{\rm cut}$ where $e_{\rm cut}$ corresponds to $r_{\rm peri}=r_{+}$. In this range of $E$, $e_{\rm cut}$ increases monotonically from 0 at $E=E_{\rm c}(r_{+})$ to 1 at $E=0$.

\label{lastpage}

\end{document}